\documentclass[12pt,preprint]{aastex}

\newcommand\MSUNYR{\rm M_{\odot}\,yr^{-1}}
\newcommand\msunyr{\rm M_{\odot}\,yr^{-1}}
\newcommand\MSUN{\rm M_{\odot}}

\newcommand\RSUN{\rm R_{\odot}}

\newcommand\Mdot{ \dot{M}}
\newcommand\mdot{ \dot{M}}
\newcommand\etal{{\rm et al}. }
\newcommand\be {\begin{equation}}
\newcommand\en{\end{equation}}
\newcommand\cm{\rm cm}

\def\cm2g{\rm cm^2 \ g^{-1}}

\slugcomment{To appear in {\it The Astrophysical Journal}}
 
\begin{document}
 
\title
{Effects of Dust Growth and Settling in T Tauri Disks}

\author{ Paola D'Alessio \altaffilmark{1}, Nuria Calvet \altaffilmark{2},  
Lee Hartmann\altaffilmark{2},  Ramiro Franco-Hern\'andez\altaffilmark{1,3},
  and Hermelinda Serv\'\i n\altaffilmark{1} }

\altaffiltext{1}{Centro de Radioastronom\'\i a y Astrof\'\i sica, 
Ap.P. 72-3 (Xangari), 58089 Morelia, Michoac\'an, M\'exico}
%\altaffiltext{2}{Harvard-Smithsonian Center for Astrophysics, 60 Garden St., 
%Cambridge, MA 02138, USA.}
\altaffiltext{2}{Dept. of Astronomy, University of Michigan, 825 Dennison Building,
500 Church St., Ann Arbor, MI 48109, USA}
\altaffiltext{3}{Harvard-Smithsonian Center for Astrophysics, 60 Garden St.,
Cambridge, MA 02138, USA.}

\email{Electronic mail: p.dalessio@astrosmo.unam.mx,  ncalvet@umich.edu,
lhartm@umich.edu, r.franco@astrosmo.unam.mx}

\begin{abstract}
We present self-consistent disk models 
for T Tauri stars which include  a parameterized treatment of dust 
settling and grain growth, building on techniques 
developed in a series of papers by D'Alessio \etal.
The models incorporate depleted distributions of dust in upper disk 
layers along with larger-sized particles near the disk midplane, 
as expected theoretically and as we suggested earlier is necessary 
to account for mm-wave emission, SEDs, scattered light images, 
and silicate emission features simultaneously.
By comparing the models with recent mid- and near-IR observations, 
we find that the dust to gas mass ratio 
of small grains at the upper layers should be $<$ 10 \% 
of the standard value.
The grains that have disappeared from the upper layers 
increase the dust to gas mass ratio of the disk interior; 
if those grains grow to maximum sizes of the order of mm
during the settling process, then both the 
millimeter-wave fluxes and spectral slopes can be consistently explained. 
Depletion and growth of grains can also enhance the ionization of upper
layers, enhancing the possibility of the magnetorotational instability
for driving disk accretion.
\end{abstract}


\section{Introduction}
\label{sec_intro}

It is  commonly accepted that disks around T Tauri stars are mostly heated 
by stellar radiation intercepted by the flared surfaces of these
disks (Kenyon \& Hartmann 1987; 1995). 
A large fraction of
the intercepted radiation is deposited in the disk upper atmosphere,
which becomes hotter than deeper layers, producing 
spectral features in emission such as the observed silicate bands
(Calvet et al. 1991, 1992; Chiang \& Goldreich 1997; D'Alessio et al. 1998;
Natta, Meyer \& Beckwith 2000).
Stellar radiation is absorbed by dust grains in the disk,
which reemit the energy at longer wavelengths. Analyses
of the spectral energy distributions
(SEDs) combined with disk structure calculations thus provide 
information on the properties and state of the solid particles in the disk.

The flat slopes of the SEDs at mm wavelengths, 
indicative of grains larger than those in the ISM,
have suggested dust evolution in T Tauri disks for some time 
(Beckwith et al. 1990; 
Beckwith \& Sargent 1991; Miyake \& Nakagawa 1993; Pollack et al 1994 = P94).
Interpretation of scattered light images in the optical and near
IR provide additional evidence for grain growth (Cotera et al. 2001; 
D'Alessio et al. 2001;
Watson \& Stapelfeldt 2004). 

In a series of papers (D'Alessio \etal 1998, 1999b, 2001; Papers I, II,
and III, respectively), we have developed
physically self-consistent models of 
T Tauri accretion disks 
and explored observational predictions based on these models.
They include both viscous dissipation and heating by irradiation
from the central star. 
 The disk structure is solved iteratively,
as the vertical disk thickness depends upon the irradiation heating, which
in turn depends upon the vertical structure.  Changes in the dust opacity
affect the vertical disk structure and therefore the temperature and density
distribution as a function of both radius and height.  

In Paper II we showed that models
in which the disk has dust like that inferred for
the diffuse interstellar medium (ISM) (i.e., silicates and graphite grains, 
with abundances and optical properties from Draine \& Lee (1984 = DL) and 
a maximum grain radius $a_{max} \sim   0.25  \ \mu m$ ) 
uniformly mixed with the disk gas  ({\it well mixed models}), 
fail to explain detailed observations.  Specifically,
the models exhibit too little mm-wave emission 
and larger far-infrared fluxes than
observed from typical CTTS for reasonable disk masses;
the models are too geometrically thick in the direction perpendicular to the
midplane, which results in wider dark lanes in scattered light images
of edge-on disks than observed; and the models predict too large a fraction
fraction of stars extincted by their disks than observed in surveys, 
assuming a random distribution of inclinations.

Motivated by the need to improve the agreement with observations,
in Paper III we considered the effects of changing dust properties
in well-mixed models.  We showed that  well mixed models 
in which dust grain radii ($a$) follow a power-law distribution 
$n(a) \propto a^{-p}$, $2.5 \leq p \leq 3.5$, with 
a maximum grain radius $\sim $ 1 mm,  
can explain the above observational constraints. 
Grains bigger than typical 
ISM dust grains affect
both the long- and short-wavelength predicted properties of the dust 
mixture; 
larger grains yield a larger
mm-wave emissivity and spectral indices in better agreement 
with observations. At the same time,
the number of small grains is reduced to account for the mass locked up
in large grains, and this decreases the optical and near-infrared opacity.
This results in
less absorption of stellar flux, lower disk temperatures, and a decreased
IR emission. Similarly, the predicted dark-lanes in scattered
light images of edge-on disks are narrower, and the expected number
of heavily extincted stars becomes consistent with observations.
However, since big grains have a gray opacity at wavelengths much shorter 
than their sizes, such disk models predict {\bf no silicate emission bands},
while these features are ubiquitous in T Tauri disks 
(Cohen \& Witteborn 1985; Natta et al. 2000), and as
strongly confirmed by large surveys
using the Infrared Spectrograph on the {\em Spitzer}
Space Telescope (Forrest et al. 2004; Furlan et al. 2005a,b).

Grain evolution is predicted by solar nebula models, in which solid
particles are found to grow and settle to the midplane, making planetesimals
and ultimately leading to planetary formation (Weidenschilling \& Cuzzi 1993,
and references therein; Weidenschilling 1997; Dullemond \& Dominik 2004b).
In such models the upper layers of the disk
are rapidly depleted of dusty material, and only
the smallest grains extend to larger vertical heights in the disk,
because the largest grains settle more quickly.
Grain growth and settling provide an explanation for the observed SEDs,
because
small grains populate the upper layers and produce conspicuous emission
dust features, but at the same time
the reduced number of small grains lowers the optical and near-infrared
opacities, while creating a population of
large particles in the disk interior consistent with high long-wavelength
emission (\S \ref{secc_overview}).

Following these suggestions,
Chiang et al. (2001) explored a two layer disk model with 
small grains
($a_{max} \sim  1 \ \mu m$) in the upper hot layer and larger 
grains ($a_{max} \sim  1$ mm) in the disk interior, 
and obtained SEDs
with conspicuous silicate and water ice bands and
high mm emission. 
In their model, the {\it irradiation surface} , that is, the surface
where the optical 
depth to the stellar radiation is unity, was taken to be  some
free parameter times the interior gas scale height.
However, if  the irradiation surface had been self-consistently calculated,
it would have been similar to the one calculated in a well mixed 
disk with small dust,
because the optical depth to the stellar radiation is built in the uppermost
layers, which in their models had a standard abundance of small grains.

More recently, Dullemond \& Dominik (2004b) followed the evolution
of dust grains as they settled in the disk and calculated the
SEDs along this evolution. They showed the main
effects of dust evolution of the SEDs, namely, that the IR fluxes
decrease as the degree of settling increases.
However, although theories of dust settling make detailed
predictions for the distribution and evolution of dust properties in the
disk, they fail to predict time scales consistent with
observations. Weidenschilling (1997) and Dullemond \& Dominik (2004b)
predicted that disks reach a large degree
of dust settling in less than 1 Myr years,
while CTTS with flared outer disks are present, in the
Taurus population, with ages of 1 - 2 Myr (Hartmann 2003)
as well as in the 10 Myr old TW Hya association (Calvet et al. 2002).
Dullemond \& Dominik (2004, 2005) suggested
several possibilities to overcome this problem. 
One suggestion was that
this might be a selection effect, but studies of the
IRAS sources in Taurus (Kenyon et al. 1990) are fairly complete
(see discussion in Paper II).  
Dullemond \& Dominik also suggested that other factors such as
different degrees of turbulence, grain properties and shapes,
uncertain sticking probabilities and other incompleteness in the
treatment of settling may explain the long lifetimes of small dust
at high disk altitudes.

In view of these problems, we take a 
parameterized
approach in this paper to describe the effects of settling. 
We parameterize the dust properties and distributions, 
and calculate the resultant
disk structure and emission, using our treatment of irradiated
accretion disks. Unlike other studies,
we obtain a surface density distribution
which is consistent with the mass accretion rate and the temperature
structure resultant from the adopted dust distribution. 
Comparison with observations will indicate the
range of parameters that best characterize populations at given
age, putting constraints into the uncertain parameters
in the physical description of grain growth and settling in
protoplanetary disks. 

Specifically,
we consider disk models with two populations of dust grains
with different spatial distributions in the disk and
different dust to gas mass ratios, small grains in the
upper levels, large grains in the midplane.
We explore the effect on the disk structure and SED 
of changing the dust to gas mass ratio in both 
populations consistently (\S \ref{sec_results}). By comparing 
to observations, specially results from the
Spitzer mission, these predictions will help to place constraints on 
the amount of dust growth
and settling (\S \ref{sec_obs_tests}).  
We close by considering the implications of our
results for dust growth models and for the ionization state
of T Tauri disks, a critical issue in determining the possible
role of the magneto-rotational instability in driving accretion
(\S \ref{sec_dis}).

\section{General considerations}

\subsection{Overview}
\label{secc_overview}

\begin{figure}
\plotone{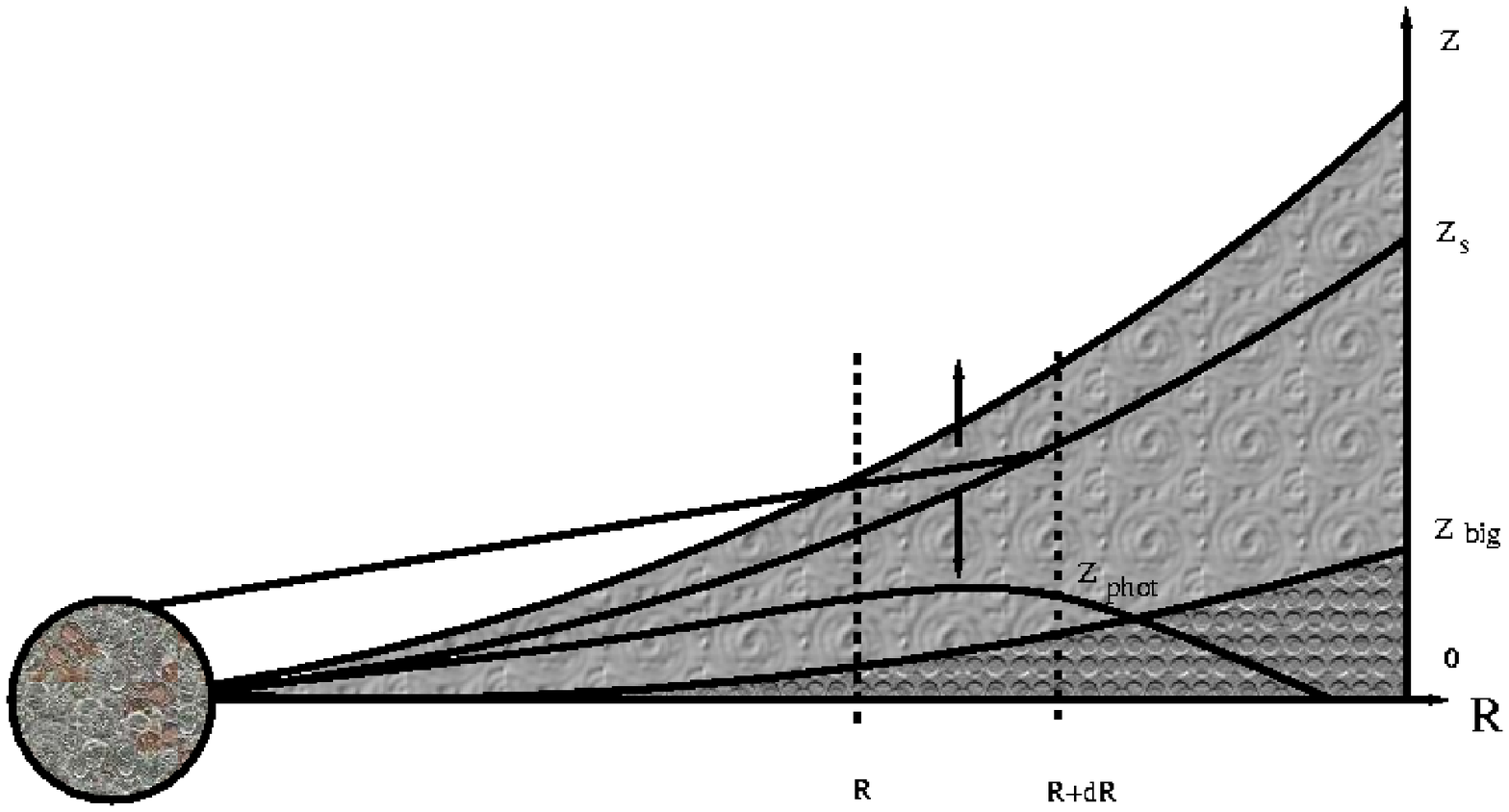}
\caption{Schematic diagram of the disk geometry. Stellar radiation is intercepted 
by the surface $z_s(R)$. The layer above $z_s$ is heated by direct stellar 
radiation, and the emission of this layer heats deeper disk regions.
The disk photosphere $z_{phot}$
 is the height where the Rosseland mean optical depth is $2/3$. 
The outer disk 
can be optically thin to its own radiation, in which case it is not possible 
to define $z_{phot}$ at those radii.
}
\label{fig_geom}
\end{figure}

The situation we are considering is illustrated
schematically in Figure \ref{fig_geom} (see also Papers I-III; 
Chiang \& Goldreich 1997).
Disk heating (at most radii) is dominated by the absorption
of light from the central star.  This radiation is absorbed
in the upper, low-density layers of the disk (the {\it absorption layers}), 
because of the
large dust opacity at the short wavelengths characteristic of the 
stellar emission and the impinging angle of this incident radiation.  
The absorbed stellar light is re-radiated at
longer wavelengths; about half of this emission enters into 
deeper layers of the disk (e.g. Chiang \& Goldreich 1997 ).  
We define an irradiation surface $z_s(R)$ such that the mean optical depth 
integrated along the radial direction to the star is unity. The amount of 
stellar flux entering the disk at a fixed $R$ is approximately 
proportional to $\mu_0$, which is the cosine of the angle between 
the radial direction and the local normal  to the surface defined by 
$z_s(R)$. Thus, the shape of 
$z_s(R)$, i.e., $\mu_0$, determines the input energy and thus 
the principal heating at most radii. 

The characteristic opacity to the stellar radiation in the upper
layers is the key agent in heating of the disk. If the deeper layers
of the disk are optically thick to their own characteristic
wavelengths of radiation, the resulting continuum emission is mostly
controlled by the amount of heating due to absorption of stellar
light, and therefore it is mostly determined by the opacity in the
upper disk layers.
The IR continuum radiation emerging from optically thick annuli 
of the disk is characteristic of the {\it photospheric height}, $z_{phot}$, 
 where the Rosseland mean vertical optical depth is $\tau_R(z) = 2/3$.
When the disk deeper layers are optically thin to their own 
characteristic wavelength of radiation (i.e., the total vertical optical depth is $\tau_R < 2/3$), 
the disk interior becomes 
vertically isothermal; it is heated by 
the stellar radiation reprocessed by the upper layers 
and it cools down by emission at longer wavelengths emerging from 
every layer.
Thus, the temperature in the outer disk, depends on both the 
short-wavelength opacity in the upper 
disk layers and the longer wavelength opacity in the deeper layers.

The emissivity as a function of wavelength of the upper optically
thin layers, heated by direct stellar radiation, depends on their dust
content.  Small grains have much higher opacities in the stellar
radiation wavelength range than large grains; thus, disks with small
grains have steeper irradiation surfaces and higher temperatures than
disks with large grains; thus, disks with small grains
have steeper irradiation surfaces and
higher temperatures than disks with large grains.  Grain growth can
be inferred not only from the slope and strength of the millimeter
flux, but from the overall level of the IR fluxes.
Using these properties, we determined that
the  median fluxes of Taurus were consistent with
a relatively low stellar absorption, as that due to large grains, 
$a_{max} =$ 1 mm, rather
than to the higher absorption of the smaller ISM dust (paper III).

Small grains will show spectral features that disappear when 
there is a significant fraction of big grains characterized by 
a more nearly gray opacity.  For instance, the $10 \ \mu m$ silicate
band is present for a dust size distribution given by 
a power law, $n(a) \sim a^{-3.5}$ between a minimum grain 
size $a_{min}=0.005 \ \mu m$ and a maximum grain size 
 $a_{max}< 1 \ \mu m$, but it tends to disappear 
for a much larger maximum size (see paper III).
Thus, the well mixed model that fitted the median SED of Taurus
could not produce the silicate emission, ubiquitous in 
T Tauri disks.

Dust growth and settling results in spatially different distributions
of grain sizes and abundances. Starting with a uniform distribution in
the disk, larger particles begin to settle toward the midplane,
leaving behind a population of small grains with a lower dust to gas
mass ratio (Weidenschilling 1997; Dullemond \& Dominik 2004b).  This
size-dependent settling has distinct effects on the spectral energy
distribution, which can be illustrated with the following approximate
description.

The emergent intensity of the optically thin upper layers is  
given approximately by (e.g., Natta, et~al.  2000), 
\be
I_\nu^{thin} \approx B_\nu(T_0) \kappa_\nu \mu_0/ \chi_*, 
\label{ithin}
\en
where $B_{\nu}$ is the Planck function at the characteristic
temperature $T_0$ of the layer, $\kappa_{\nu}$ is the dust true
absorption at frequency $\nu$, and $\chi_*$ is the Planck mean opacity
of the dust evaluated at the stellar effective temperature $T_*$
(calculated integrating the monochromatic total opacity over frequency
using $B_\nu(T_*)$ as the weighting function).
The temperature $T_0$ at radius $R$ can be estimated 
assuming that the dust is in radiative equilibrium with the 
stellar radiation field, i.e.,  

\begin{equation}
T_0 (R) \approx 0.8 [\kappa_*/\kappa_P(T_0)]^{1/4} (R_*/R)^{1/2} T_*\,,
\label{tthin}
\end{equation}
where $\kappa_*$ and $\kappa_P(T_0)$ are Planck mean true absorptions
evaluated at the stellar effective temperature and the local
characteristic temperature, respectively,
and $R_*$ is the stellar radius.
The factor 0.8 in equation (\ref{tthin}) accounts for the fact that most
of the emission of the
layer emerges from a depth close to $z_s$ (see Fig. 1), where the incident 
stellar radiation that heats the dust has been attenuated.

The approximate equations (\ref{ithin}) and (\ref{tthin}) show that the
temperature and spectrum of the upper layers will change
with the wavelength dependence of the opacity but
not with the dust to gas mass ratio, since they both
depend on opacity ratios [namely $\kappa_\nu/\chi_*$ and
$\kappa_*/\kappa_P(T_0)$]. 
On the other hand, lowering the dust to gas mass ratio
in the upper layers, decreases their opacity at the
stellar radiation,  making $z_s(R)$ flatter and $\mu_0$ smaller.
These considerations suggest that a model with small grains at the 
upper layers, with a smaller dust to gas mass ratio in those layers
such that the disk roughly has  the same surface
$z_s(R)$ than the well mixed model with $a_{max} =$ 1 mm 
should be 
able to explain both, the median SED and the 
observed 10 $\mu m$-silicate band in emission.
More generally, the observed SEDs and, in particular, 
the mid-IR spectral range, can be used to quantify 
the spatial distribution of the dust in the disks around CTTS, and thus
the degree of dust settling toward the midplane.

\subsection{Settled disk Models}
\label{secc_models}

We construct self-consistent viscous steady irradiated disk models
with dust settling. To describe the disk turbulent viscosity, we use 
the $\alpha$ prescription (Shakura \& Sunyaev 1973), i.e., the viscosity
coefficient is taken to be $\nu_t = \alpha c_s H$, where $c_s$ and $H$ are 
the local sound speed and gas scale height, respectively, and $\alpha$ 
is a free parameter, assumed to be constant throughout the disk. These are  
irradiated $\alpha$-disk models. 

We have 
included dust settling in a simple way.
For each dust ingredient, we adopt a grain size distribution 
given by a power law of the grain 
radius, $n(a)= n_0 \ \ a^{-p}$, between a minimum and maximum radius 
($a_{min}$ and $a_{max}$), with an exponent $p$. 
 The constant $n_0$ is related to the 
dust to gas mass ratio $\zeta$ of the particular ingredient, 
the grain bulk density, the minimum and maximum sizes,   
and  $p$ (see Appendix).
We explore {\it settled} disk models with  two populations of grains
with different size distributions.  
We describe in detail models in which the 
{\it big grains} component has  $a_{min}=0.005 \ \mu m$, 
$a_{max}=$ 1 mm, $p=3.5$,  and the {\it small grains} component, 
 $a_{min}=0.005 \ \mu m$, $a_{max}=0.25 \ \mu m$, $p=3.5$,
and in \S \ref{secc_amax} we discuss 
the effect of changing the small grains maximum 
size.
 Both populations have different spatial distributions and dust to gas 
mass ratios, $\zeta$, i.e., 
the big grains are concentrated close to the midplane and $\zeta_{big}$ 
may be larger than the standard value because this population incorporates
 the mass of the dust 
that has settled from upper layers;  the small 
grains are well mixed with the gas in the rest of the disk, having a lower 
$\zeta_{small}$ than the standard value (i.e., the small grains are depleted). 
This is an attempt to model in a simple way the complex outcome from 
detailed studies of the growth of disk dust grains.

In these exploratory calculations, we reduce the parameter space 
by considering  that the big grains are concentrated in the midplane, 
below a  fixed height $z_{big} =0.1 \ H$, where $H$ is the local gas 
scale height.  The precise value of $z_{big}$ is arbitrary; we 
make it small with respect to $H$ to simplify matters.
As shown schematically 
in Figure \ref{fig_geom}, the small value of $z_{big}$ assures 
that the upper layer that absorbs the 
stellar radiation is in the region with the small grains population.
To simplify even more, we assume that 
the grain size distribution and dust to gas mass ratio 
do not change with radius, but only with height, and 
we take the values of the grain maximum size in both populations 
to be spatially constant. 
We decrease the free parameter $\zeta_{small}$ and
increase $\zeta_{big}$  
to keep the total dust-to-gas mass ratio at the initial
(solar) value.  Thus we ignore any possible radial concentrations
or depletions of solids (e.g., Youdin \& Shu 2002).

The calculation of $\zeta_{big}$ as a function of $\zeta_{small}$ 
is described in the Appendix.
We quantify the depletion of the small grains  
relative to the standard dust to gas mass ratio using the parameter  
\be
\epsilon=\zeta_{small}/\zeta_{std}
\en
where $\zeta_{std}$ is the standard
(initial) dust to gas mass ratio. 
In \S \ref{secc_zbig} we describe the effects of changing $z_{big}$.

\begin{figure}
\plotone{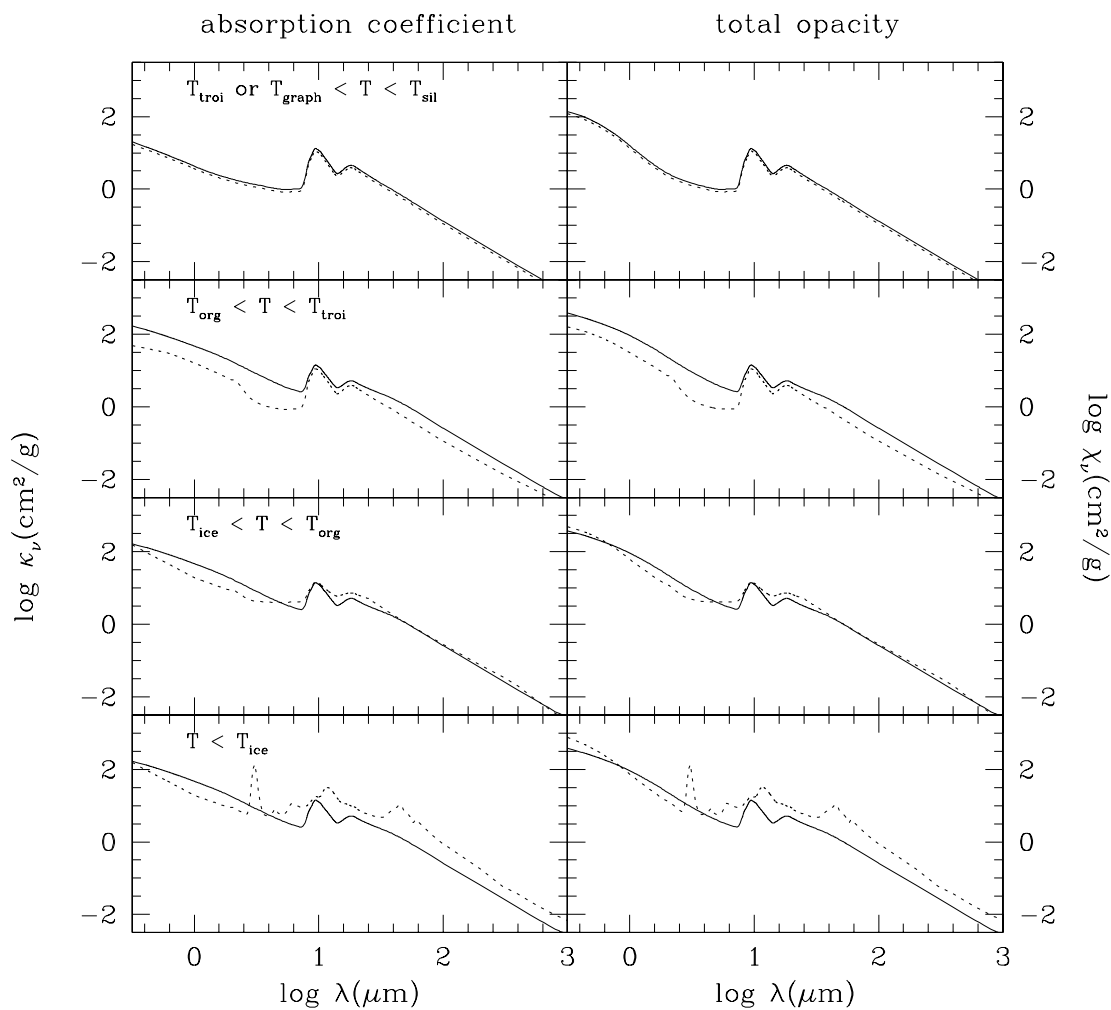}
\caption{Opacity of the small grains in the upper layer. 
The left panels show the absorption coefficient 
and the right panels show the total opacity (absorption + scattering). 
We compare coefficients for the ISM mixture (solid line) and 
the P94 mixture (dotted line), for different ranges of temperature
defined using the sublimation temperatures listed in Table \ref{table_dust}}
\label{fig_opas}
\end{figure}

An important ingredient of the present models 
 is the dust composition and optical properties.
We explore two main compositions, one consistent with the disk dust model
proposed by P94 and the other one 
corresponding to the model proposed by DL 
for the diffuse ISM, to show the effects of
dust content. 
In a future paper, we will attempt detailed fits to individual
spectra to explore the range of dust compositions present in young
disks.
Table \ref{table_dust} lists the adopted 
ingredients with their
standard fractional abundances, $\zeta_{std}$, and sublimation temperatures.
For simplicity, we ignore the dependence of sublimation temperature on 
gas density and grain size. 
Figure \ref {fig_opas} shows the monochromatic 
absorption coefficient or 
true opacity and the total opacity for the dust mixtures adopted 
and different temperature ranges.

%\clearpage
\begin{deluxetable}{cllr}
\tablecaption{Dust ingredients\label{table_dust}}
%\tablewidth{50pt}
\tablehead{
\colhead{composition}  & \colhead{ingredient} &\colhead{$\zeta_{std}$} 
&\colhead{$T_{sub}(K)$}}
\startdata
ISM & silicates &0.004  &1400\\
&  graphite & 0.0025 &1200\\
P94 & silicates &0.0034 &1400  \\
& organics & 0.0041 & 425\\
& troilite & 0.0008 & 680 \\
& water ice & 0.0056 & 180
\enddata
\end{deluxetable}
%\clearpage

Given the dust composition, the maximum 
grain sizes and the dust to gas mass ratios, the irradiated 
accretion disk model is calculated as described in papers I, II and III. 
The input parameters for the disk structure calculations are the mass, radius and effective 
temperature of the central star ($M_*, \ R_*$ and $T_*$), 
the disk mass accretion rate $\Mdot$, and the viscosity parameter $\alpha$, 
assumed both to be constant as a function of distance to the central star,
as well as the settling parameter $\epsilon$.
The disk structure equations in the vertical direction are solved
using an iterative scheme 
which recalculates the absorption of stellar radiation 
self-consistently given the disk structure in each step.

Two additional parameters have to be specified to obtain the SED: 
the inclination angle
of the disk axis relative to the line of sight, $i$,
and the disk maximum radius, $R_d$.
The SED is calculated by integrating 
the transfer equation, including thermal and scattering emissivities,
 along rays 
that cover the disk projection in 
the plane of the sky  for a given inclination angle 
 and for each wavelength.

We assume the gas and the dust are in thermal
balance even at the uppermost layers, which is a good approximation
close to $z_s$ (e.g., Kamp \& Dullemond 2004) and is correct
at deeper denser layers.
This assumption implies
that a unique temperature characterizes not only gas and dust, but also each
grain, regardless its size or composition.
For the temperatures of the disk models, hydrogen is in molecular form
except in the midplane of the innermost regions of the high mass
accretion rate disks. The adopted assumption breaks down for heights
above the midplane larger than $z_s$.  In these low density layers,
collisions between gas and dust cannot equalize temperatures of gas
and dust (Glassgold et al. 2004) nor temperatures of different grains
(Chiang et al. 2001). In particular, the gas temperature in the
uppermost levels becomes high enough to dissociate molecules (Aikawa
\& Herbst 2001; Aikawa et al. 2002; Bergin et al. 2003; Willacy et
al. 1998; Willacy \& Langer 2000).  These effects are essential for
analysis of molecular observations or detailed studies of dust
features, but are less important for the bulk of dust thermal emission
considered in this paper.

We show results for a fiducial set of models, which have
typical parameters for the CTTS in Taurus,
$M_*=0.5 \ \MSUN$, $R_*=2 \ \RSUN$, $T_*=4000$ K, $\Mdot=10^{-8} \ \MSUNYR$, 
$\alpha=0.01$, $R_d=300$ AU and $i=50^\circ$.
The inclination angle $i=50^\circ$ (or $\cos \ i=0.65$) is adopted 
instead of the typical angle $i=60  $ ($\cos \ i=0.5$),
which is the average of
$\cos i$ for random disk inclinations; this is an attempt 
to account for the fact that highly inclined disks
would have peculiar  SEDs 
and would not be  classified as
a CTTS SED (see papers II and III; Chiang \& Goldreich 1999; \S \ref{secc_angles}).
We explore in detail the dependence of the structure and emission
on $\epsilon$ for models with these parameters; in \S
\ref{secc_mdots} and \ref{secc_angles}, we study the dependence of the SED
on $\mdot$ and inclination.

%We assume the gas and the dust are in thermal
%balance even at the uppermost layers, which is a good approximation
%close to $z_s$ (e.g., Kamp \& Dullemond 2004) and is correct
%at deeper layers.
%This assumption implies
%that a unique temperature characterizes not only gas and dust, but also each
%grain, regardless its size or composition.
%%--------
%Given the low temperatures and high densities, hydrogen is
%mostly molecular in most of the disk (except in the midplane
%of the innermost disk regions in the high mass accretion rate
%disks). The adopted assumption
%breaks down for heights above the midplane higher than $z_s$.
%With the low densities of these layers, collisions cannot
%keep gas and dust in thermal balance, and the gas temperature
%becomes higher than the dust temperature due to irradiation
%by X-rays and UV radiation (Glassgold et al. 2004), resulting
%molecular dissociation in the uppermost layers. These effects
%must be included when analyzing molecular observations of
%disk, but are of much less important for dust thermal emission.

The disk models include the emission from the wall at the dust sublimation radius
(Natta et al. 2000; Tuthill et al. 2001; Dullemond, Dominik \& Natta 2001; 
Muzerolle et al. 2003). 
In our models,
both the central star and the accretion shocks at the
stellar surface heat the dusty wall (Muzerolle et al. 2003; D'Alessio et al. 2004). 
Assuming that 
the dust population at this radius only consists of
silicates with the {\it small} grain size distribution,
using the silicate sublimation temperature in Table \ref{table_dust},
this radius is $R_{sub}=10 \ R_*$ for 
the fiducial parameters.
We take this value as the minimum disk radius, 
since for the range of mass accretion rates and low
stellar temperatures considered the gaseous inner disk is 
probably optically thin, so it does
not shadow the wall from stellar radiation  and, compared to the 
irradiated wall, does not contribute appreciably to the SED (Muzerolle et al. 2004).

For consistency, we generally assume that the dust in the wall is settled
in the same proportion as the rest of the disk. This 
assumption affects the estimated height of the wall, which is calculated 
as the height where the optical depth to the stellar radiation is unity  
(Dullemond et~al. 2001; Muzerolle et al. 2004). 
Thus, the emergent flux of a wall corresponding to
a disk with $\epsilon <1$ is smaller (because the wall observed area 
is smaller) than in the $\epsilon=1$ case.

The treatment described in papers I, II and 
III corresponds to a disk in which the scattering of stellar radiation is 
{\it isotropic}. However, depending on their composition and 
size,  dust grains tend to scatter light in a more forward direction. 
The extreme case of perfectly forward scattering should give 
a structure and SED indistinguishable from a case where only true absorption 
is being considered (Natta et~al. 2000). 
This is the case were the intensity emerging from the upper optically 
thin layer would be maximum, since the stellar radiation can 
penetrate and heat the deepest.
So, in addition to our calculations with isotropic scattering, we present
results for the limiting case of completely forward scattering (i.e., no scattering)
to quantify the consequences of these effects.

\section{Results}
\label{sec_results}

\subsection{Effects of dust depletion on the disk structure}
\label{secc_struc}

In this section we describe the effect of dust settling toward
the midplane characterized as described in \S \ref{secc_models} 
on the disk structure. 
We compare the results for disk models with settling with 
those for the well mixed disk model with $a_{max}=$ 1 mm
that approximately fits the median SED of 
Classical T Tauri Stars in Taurus (paper III).
Table \ref{table_prop} lists different properties of the models discussed here.

%Note that the mass of the models is not fixed. The mass is 
%the integral of the mass surface density over radius, and thus depends on
%radius. In addition, since the surface density depends on the temperature
%structure
 
\begin{figure}
\plotone{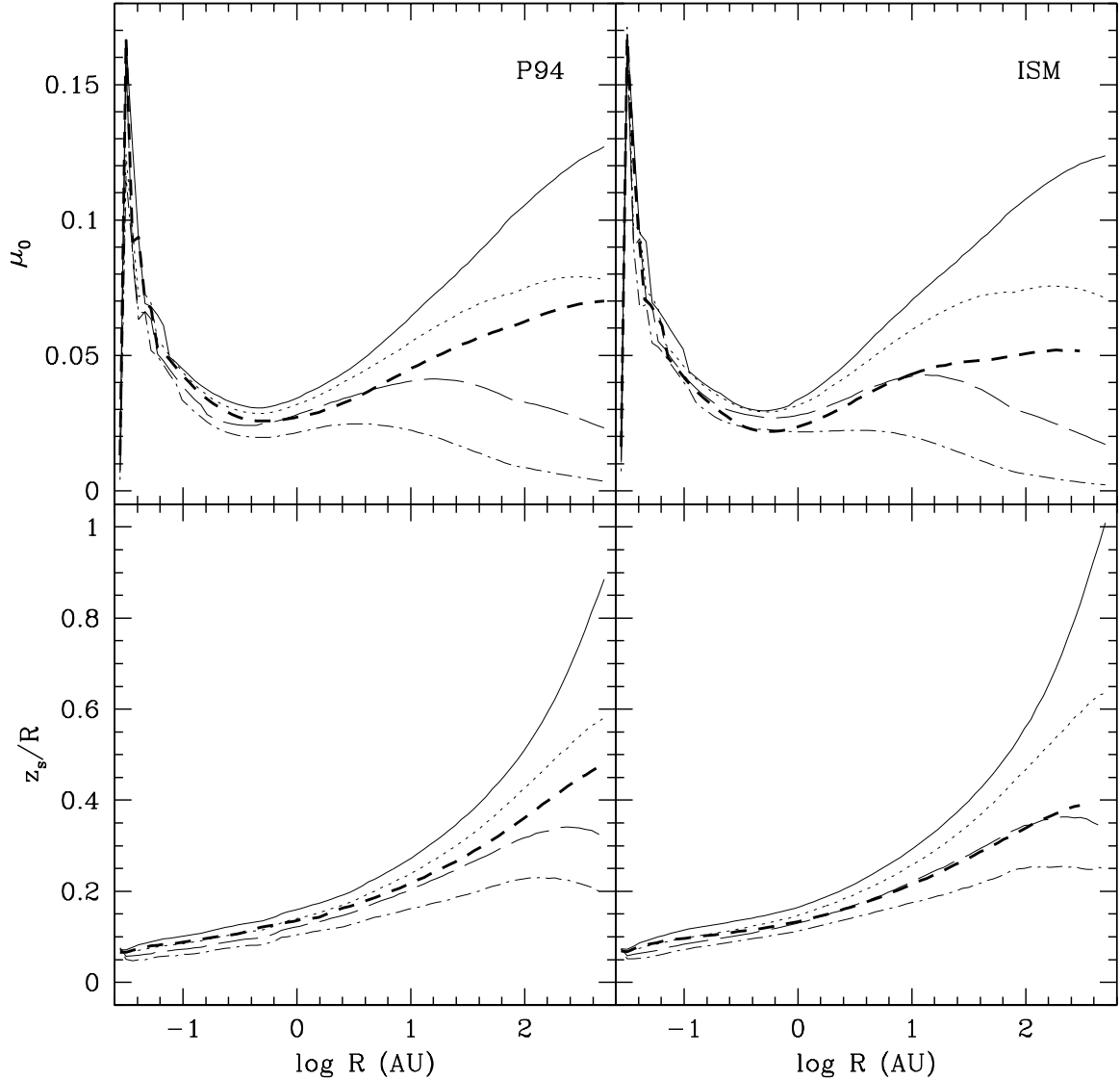}
\caption{Upper panels: Cosine of the angle between the impinging 
stellar radiation and the normal to the disk surface.
Lower panels:
Height of the irradiation surface divided by radius, $z_s/R$, in settled 
disk models. In each case, the left panel corresponds to P94 dust 
and the right panel to ISM dust. 
Each curve corresponds to a different depletion of the small grain component, 
$\epsilon$ = 1 (solid line), 0.1 (dotted line), 0.01 (long-dashed line), 0.001 (dot-dashed line). The heavy
dashed line corresponds to a well mixed model with $a_{max}=$ 1 mm.
A model with only 
small grains and no depletion, 
is indistinguishable from the case with no depletion, $\epsilon=1$,  
}
\label{fig_superf}
\end{figure}

The direct consequence of decreasing $\zeta_{small}$ is to decrease
the opacity to the stellar
radiation and thus lower the height of the irradiation surface
$z_s(R)$.
Figure \ref{fig_superf} shows 
how $z_s(R)$ and $\mu_0(R)$ (cf. \S \ref{secc_overview}) decrease 
when $\zeta_{small}$ and $\epsilon$ decrease.
The change in $\zeta_{small}$ (or $\mu_0$) affects the irradiation flux, since 
the smaller the $\mu_0$ the smaller the stellar flux intercepted by the disk 
and the colder the disk. 
We see that the well mixed model with $a_{max}=$ 1 mm has a $\mu_0$ 
between the models with  $\epsilon = 0.1$ and $\epsilon=0.01$.
The irradiation surface is slightly higher in the case of ISM dust,
because of the higher opacity of this mixture at the wavelength
range where the stellar radiation is mostly absorbed (see Figure \ref{fig_opas}).

For the fiducial models shown in 
Figure \ref{fig_superf}, the curvature of the disk surface is not important 
for $R < 0.2$ AU, and  $\mu_0(R)$ is similar 
to that of  a flat disk, 
$\mu_0 \approx {2 / 3 \pi}(R/R_*)^{-1}$.
However, at larger radii $\mu_0(R)$ increases 
with radius, and even for $\epsilon=0.001$ (i.e., with 0.1 \% of the 
standard dust to gas mass ratio in the absorption layers) 
 $\mu_0(R)$ is larger than for a perfectly flat disk.
As Dullemond \& Dominik (2004b), we also find that for very low values of 
$\epsilon$ (i.e., a more advanced stage of settling), $z_s/R$ has a maximum 
(at $\sim$ 100 AU) and decreases for larger radii. This type of geometry 
led Dullemond \& Dominik to argue that the outer disk of highly settled models
is in the shadow of inner regions and does not receive stellar radiation. 
However, the scattering of stellar 
radiation by dust above $z_s$ and the radial transport of energy would 
tend to increase the heating of the outer disk and remove 
the shadowing effect. Since our model is 1+1D 
we cannot include the scattering in the uppermost 
layers and the radial transport of energy. 
In any case, the contribution of disk regions beyond 100 AU to the SED 
for highly settled models
is negligible in the mid-IR, which is the spectral region where the 
effects of settling are most important (cf. \S \ref{secc_spatial}.)  

\begin{figure}
\plotone{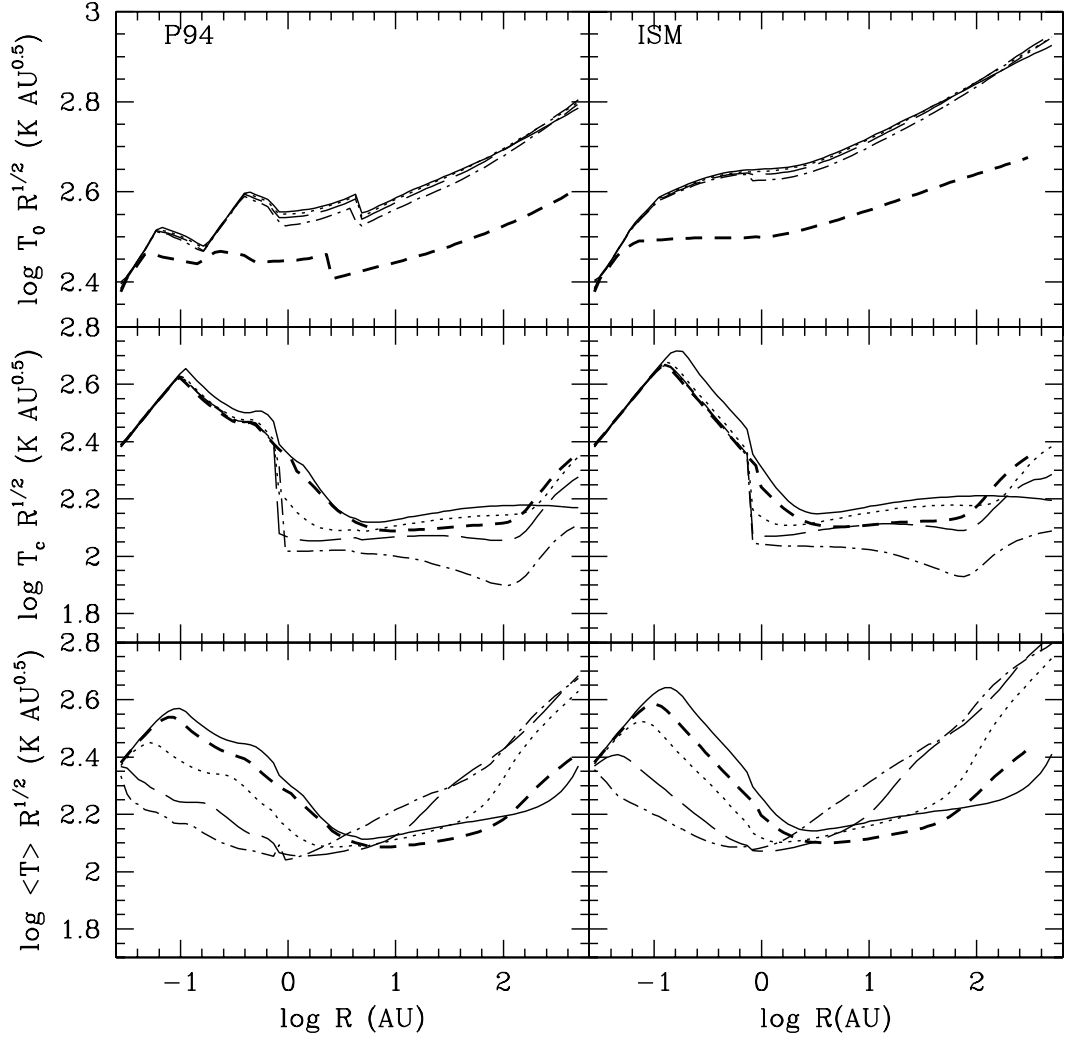}
\caption{ Characteristic temperatures in the disk. Upper panels: Surface temperature. 
Middle panels: Midplane temperature. Lower panels: Mean temperature 
weighted by density. To emphasize differences we plot 
log $(T \times  R^{1/2})$. Models are shown for two dust composition
and four values of $\epsilon$ 
(line types as in Fig. \ref{fig_superf}).
}
\label{fig_temperaturas}
\end{figure}

Figure \ref{fig_temperaturas} shows different characteristic temperatures 
of the disk as a function of radius.
The upper panels show the surface temperature $T_0$ for different values of
dust depletion in the upper layers $\epsilon$  and the two adopted dust compositions. 
To show the difference between models in an expanded format we have plotted 
log $(T_0 \times  R^{1/2})$, because the temperature distribution
tends to approximately $T_0 \propto  R^{1/2}$ at large radii.
One can see that all the models with small grains in the upper atmosphere
and the same dust composition have similar $T_0(R)$, regardless the 
value of $\epsilon$; this temperature is similar to that
of the well mixed model with small grains.
This is consistent with what one expects from radiative equilibrium of 
optically thin dust, in which the heating and the cooling are both 
proportional to the dust to gas mass ratio, and then the dependence 
in $\zeta_{small}$ (or $\epsilon$) cancels out (\S \ref{secc_overview}).
The small differences are due to the inclusion of the heating of the 
upper dust grains by radiation emerging from the disk own photosphere 
(see Calvet \etal 1991).

The cases with P94 dust show discontinuities which are related to the sublimation 
temperatures of the different ingredients (see Table \ref{table_dust}). 
Comparing temperatures for the ISM and P94 mixtures,
one can see that the presence of water ice in the outer disk increases the cooling.
For instance, at 100 AU the surface temperature for the models with 
P94 dust (with water ice) is a factor of 1.4 lower than 
the surface temperature of the models with ISM dust (without 
water ice).

The surface temperature of the settled models in
Fig. \ref{fig_temperaturas} is higher than that of 
the well mixed model with big grains ($a_{max} = 1$ mm),
because small grains are much more efficient absorbing stellar radiation.
However, even for $a_{max}=$ 1 mm, $T_0 \times R^{1/2}$ is not constant as 
expected if grains were blackbodies (i.e., with gray opacities); 
since we have a grain size 
distribution such that there is always a fraction of small grains 
(with $a < \lambda/2 \pi$) regardless how big $a_{max}$ is, so that 
the resulting opacity is not completely gray.

The mid panels of Figure \ref{fig_temperaturas} show the midplane temperature 
as a function of radius. Again, to emphasize the differences between models
we plotted log $(T_c \times  R^{1/2})$ in the vertical axis, instead of 
log $\ T_c$.  
In contrast to the surface temperature, the midplane temperatures show 
a strong dependence on $\epsilon$, decreasing as $\epsilon$ decrease and less
stellar energy is captured.
The cases with $\epsilon=0.01$ and $0.001$, have lower temperatures 
than a well mixed model with $a_{max}=$1 mm for $R >$ 10 AU, as a consequence of
the lower irradiation surface of these settled models (Fig. \ref{fig_superf}). 

\begin{figure}
\plotone{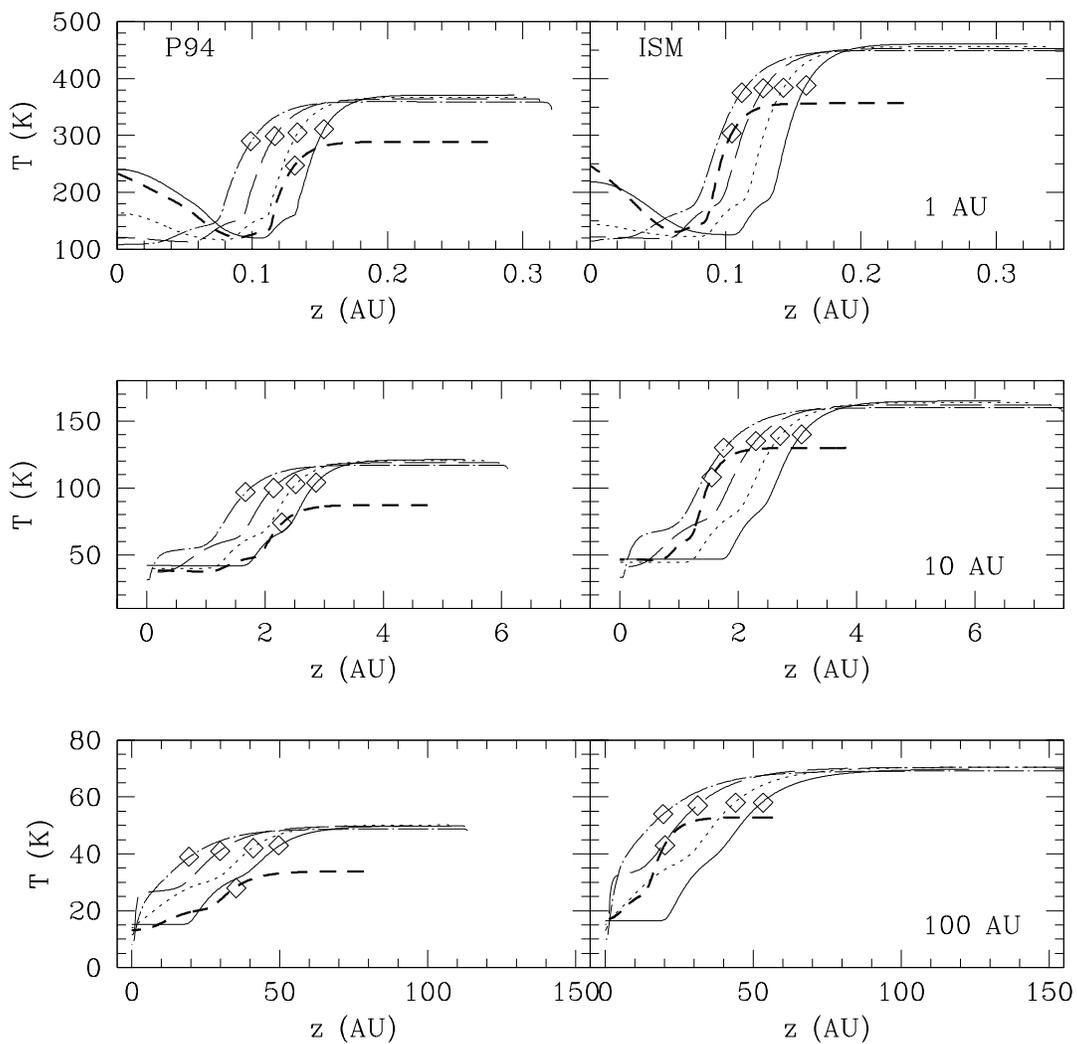}
\caption{Temperature vs height at different radius. Upper panels: $R=$ 1 AU. 
Middle panels: $R=$ 10 AU. 
Lower panels:  $R=$ 100 AU.
The diamonds mark the height of the irradiation
surface, $z_s$. 
Models are shown for two dust composition
and four values of $\epsilon$ 
(line types as in Fig. \ref{fig_superf}).
}
\label{fig_vertical}
\end{figure}

To see in more detail the dependence of the temperature on $\epsilon$,
Figure \ref{fig_vertical} shows the vertical distribution of temperature
at three representative radii: 1, 10 and 100 AU, for both
dust compositions and different depletion factors.
For comparison we also show the
vertical temperature distribution for the well mixed model 
with $a_{max}=$ 1 mm.
The behavior of $T(z)$ for well mixed models has been
discussed before (Calvet et al. 1991, 1992; Paper II).
In short, the temperature has a plateau at $\sim T_0$ at large heights
above the midplane,
which roughly corresponds to the absorption layer discussed in \S \ref{secc_overview}.
Then, at a height of $\sim$ 1.3 - 1.4 $z_s$ ($z_s$ is indicated in the Figure),
the temperature begins to decrease as $z$ decreases, since less stellar
energy reaches those layers. Near the midplane at radii in
where the disk is thick to its own radiation (cf. R = 1 AU in Figure \ref{fig_vertical}), 
the temperature
gradient becomes positive, so that the viscous flux generated
near the midplane is transported outward. When the
disk becomes optically thin to its 
own radiation at larger radii
(cf. R = 10 AU, 100 AU in Figure \ref{fig_vertical}), 
the midplane becomes nearly isothermal.

As shown in Figure \ref{fig_vertical},
as $\epsilon$ decreases and the upper layers become more
depleted of grains, stellar radiation penetrates deeper into the disk
changing the vertical temperature structure.
The upper plateau at $T_0$ is still present, but
it extends deeper into the disk as the height $z_s$
decreases with $\epsilon$.
In the dense midplane of the inner regions,
temperatures drop because the Rosseland mean optical depth
decreases. In the outer regions, 
as the bulk of the disk becomes
more and more optically thin, an increasing amount of stellar
radiation, reprocessed by the absorption layers,
reaches regions closer to the midplane.
$T(z)$ has a further drop
in the big grains layer near the midplane;
as $\epsilon$ decreases, more mass concentrates
in this layer, the optical depth
of this layer increases, and less energy
reaches the level $z=0$, lowering $T_c$ 
(cf. Figure \ref{fig_temperaturas}).

\begin{figure}
\plotone{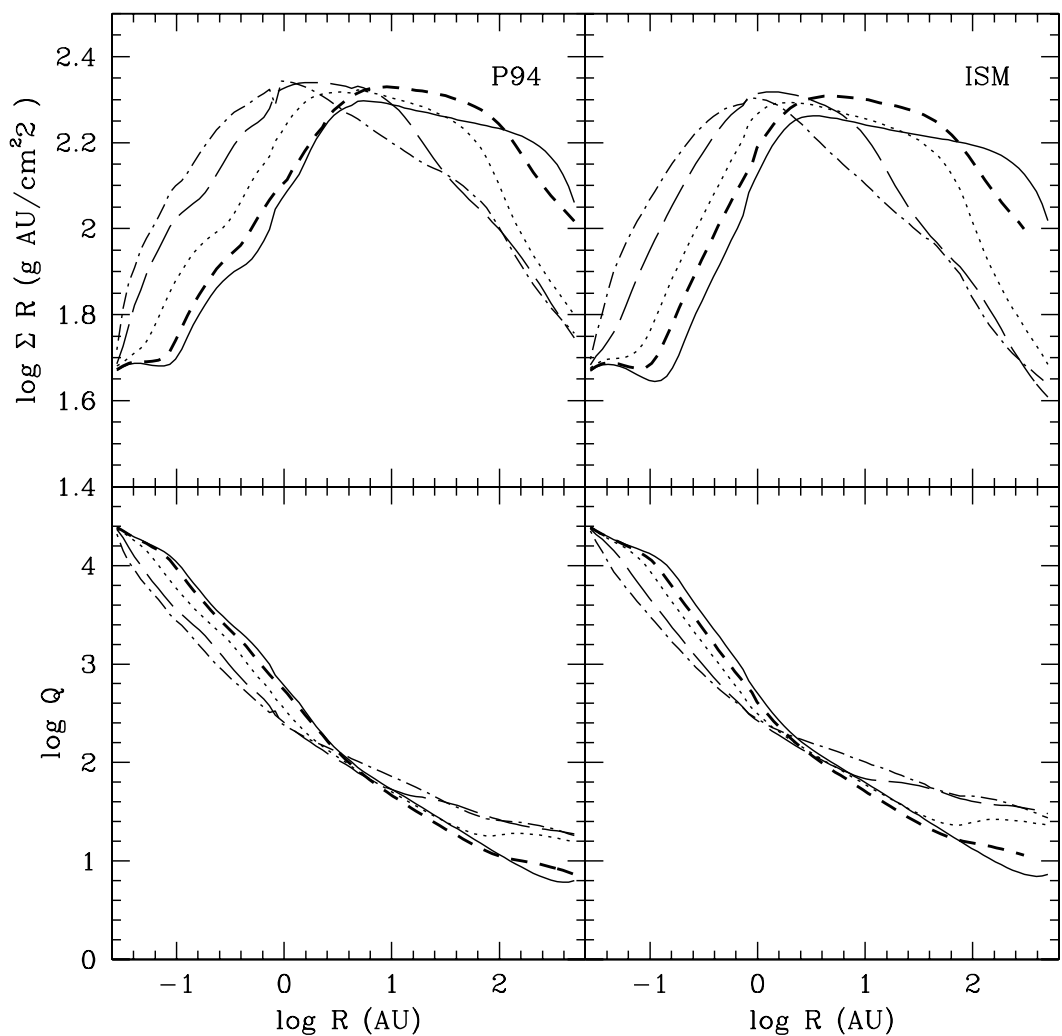}
\caption{Characteristic disk properties.
Upper panels: Mass surface density times radius as a function 
of radius.  Note that the true surface density does not
decline towards the central star.
%Each curve (solid line) corresponds to a settled model 
%with a different value of $\epsilon$ from 1 to 0.001 (density 
%increases for R < 1 AU when $\epsilon$ decreases). 
%We also plot the surface density of well mixed models with $a_{max}=0.25 \ \mu$m (dotted line) and 1 mm (dashed line).  
Lower panels: Toomre parameter evaluated at the disk midplane 
for the same models. A Toomre parameter $Q \lesssim 1$ indicates that
the disk is gravitationally unstable.
Models are shown for two dust composition
and four values of $\epsilon$ 
(line types as in Fig. \ref{fig_superf}).
}
\label{fig_sigma}
\end{figure}

The models presented here are $\alpha$ steady accretion disk models,
for which the mass surface density is consistently obtained in terms
of the mass accretion rate $\Mdot$ and 
the viscosity, which in the 
$\alpha$ prescription depends on temperature.
Since the temperature structure depends on $\epsilon$,
so does the surface density and the disk mass, for the
same mass accretion rate.
Figure \ref{fig_sigma} shows $\Sigma \times  R$ for
models with different depletions (again, to compare
with the standard assumption $\Sigma \propto 1/R$).
The well mixed model with $a_{max} =$ 1 mm is also
shown for comparison. 

To understand the behavior
of $\Sigma$, note that
with the adopted assumptions, 
conservation of angular momentum flux gives
$\int_0^\infty \rho \nu_t dz = \Mdot/3 \pi (1-(R_*/R)^{1/2})$,
where the viscosity is 
$\nu_t(z) = c_s / H = \alpha c_s^2(z) /\Omega_k$,
with $H = c_s(z)/ \Omega_K$, where
$c_s(z)$ is the sound speed at $z$, and
$\Omega_K$ is the
Keplerian angular velocity.
This expression
is usually approximated in well mixed models as
$\Sigma \sim \Mdot \Omega_k/3 \pi \alpha c_s^2(T_m) (1-(R_*/R)^{1/2}) $ 
with $T_m \sim T_c$,
because $T_c$ characterizes the region near the midplane
where most of the mass is. 
However, because we have adopted a $z_{big}$ considerably less
than $H$, most of the disk mass (which is contained within $H$)
is {\em not} characterized by $T_c$.
In this cases, $T_m = <T>$, where $<T>$ is the average
temperature weighted by the local volumetric mass density;
this temperature is shown in the lower panels of Figure \ref{fig_temperaturas}. 
The surface density follows the inverse of the average temperature.

At small radii, temperatures within one scale height of the midplane
decrease with $\epsilon$ (see Fig. \ref{fig_vertical}),
so the average temperature decreases and $\Sigma$ increases.
At large radii, the temperature is low at the midplane, but
begins to increase in less than one scale height, so that
$<T>$ increases with $\epsilon$ and $\Sigma$ decreases.

The dependence of $\Sigma$ on radius 
can be seen in Figure \ref{fig_sigma}. For $\epsilon$ =1,
approximately $\Sigma \propto R^{-1}$ at radii from a few to $\sim$ 200 AU.
As $\epsilon$ decreases $\Sigma$ decreases more steeply, with
$\Sigma \propto R^{-1.2}$ for $\epsilon$ = 0.001.
In all cases, $\Sigma \propto R^{-1.4}$ at large radii where the disk 
becomes optically thin.
The corresponding disk masses, calculated from the integration of the
surface density, are given in Table \ref{table_prop} for disk radii of
100 AU and 300 AU.  For the same mass accretion rate, disks with no
depletion are $\sim$ 2 more massive than those with $\epsilon$ = 0.001
(0.1 \% depletion), because of the lower surface density of the
latter.
The specific angular momentum of the disk is given by

\be
{J_d  \over M_d}= 4 \pi \int_{R_{mag}}^{R_d} R^3 \Sigma(R) \Omega_K(R) dR
/ M_d
\en
where $\Omega_K$ is the Keplerian angular velocity.  Using as a
reference the angular momentum of an annulus with the disk mass
concentrated at the outermost radius, i.e., $(J_d/M_d)^0= (G M_*
R_d)^{1/2} = 3.15 \times 10^{20} (R_d/100 AU)^{1/2} cm^{2} s^{-1}$, we find
that all the models displayed in Table 2 have $J_d/M_d \sim 0.6-0.7
(J_d/M_d)^0$.

Given the changes in the temperature and surface density relative to
the typical well mixed models with the same mass accretion rate, it is 
important to quantify the Toomre parameter of these models to check 
their stability against self-gravity perturbations. 
The lower panels of Figure \ref{fig_sigma} show the Toomre parameter evaluated 
at the disk midplane, 

\begin{equation}\label{QDEF}
Q \equiv \frac{c_s (T_c)\Omega}{\pi G \Sigma}. 
\end{equation}
We can see that the stability condition $Q \lesssim 1$ 
(e.g., Gammie \& Johnson 2004) 
is satisfied at every radius for the present set of parameters.

%\clearpage
\begin{deluxetable}{cccccc}
\tablecaption{Model properties \label{table_prop}}
%\tablewidth{50pt}
\tablehead{
\colhead{$\epsilon=\zeta_{small}/\zeta_{std}$} & \colhead{Dust}  &\colhead{$w$} &\colhead{$M_d(100 \ AU)/\MSUN$}  &\colhead{$M_d(300 \ AU)/\MSUN$} &\colhead{$z_{wall}(R_*)$ } } 
\startdata
1 & ISM & $\ne 0$  &$1.2 \times 10^{-2}$  &$3.2 \times 10^{-2}$ &$1.41$\\
0.1 & " & $\ne 0$ &$1.1 \times 10^{-2}$  &$2.0 \times 10^{-2}$ &$1.23$\\
0.01 & " & $\ne 0$ &$7.8 \times 10^{-3}$  &$1.5 \times 10^{-2}$ &$1.0$  \\
0.001 & " & $\ne 0$ &$7.1 \times 10^{-3}$  &$1.4 \times 10^{-2}$ &$0.76$  \\
1 & ISM & 0 & $1.1 \times 10^{-2}$ &$3.2 \times 10^{-2}$ &$1.12$\\
0.1 & "  &0 &$1.1 \times 10^{-2}$  &$2.2 \times 10^{-2}$ &$0.95$\\
0.01 & "  &0 &$8.9 \times 10^{-3}$  &$1.6 \times 10^{-2}$ &$0.75$ \\
0.001 & "  &0 &$7.3 \times 10^{-3}$  &$1.5 \times 10^{-2}$ &$0.47$ \\
1 & P94 & $\ne 0$ &$1.4 \times 10^{-2}$  &$3.5 \times 10^{-2}$ &$1.41$ \\
0.1 & "  & $\ne 0$&$1.3 \times 10^{-2}$  &$2.4 \times 10^{-2}$ &$1.23$ \\
0.01 & " & $\ne 0$ &$9.9 \times 10^{-3}$  &$2.0 \times 10^{-2}$ &$1.0$ \\
0.001 & " & $\ne 0$ &$9.7 \times 10^{-3}$  &$2.0 \times 10^{-2}$ &$0.76$  \\
1 & P94  &0 &$1.1 \times 10^{-2}$  &$3.3 \times 10^{-2}$ &$1.12$\\
0.1 & "  &0  &$1.3 \times 10^{-2}$  &$2.7 \times 10^{-2}$ &$0.95$\\
0.01 & "  &0&$1.1 \times 10^{-2}$  &$2.2 \times 10^{-2}$ &$0.75$  \\
0.001 & " &0 &$9.7 \times 10^{-3}$  &$2.0 \times 10^{-2}$ &$0.47$ \\
\enddata
\tablenotetext{~}{Notes: $w$ is the albedo. $M_d$ is the disk mass up
to the radius inside the parenthesis.}
\end{deluxetable}
%\clearpage

\subsection{Spectral Energy Distributions}
\label{secc_emission}

\begin{figure}
\plotone{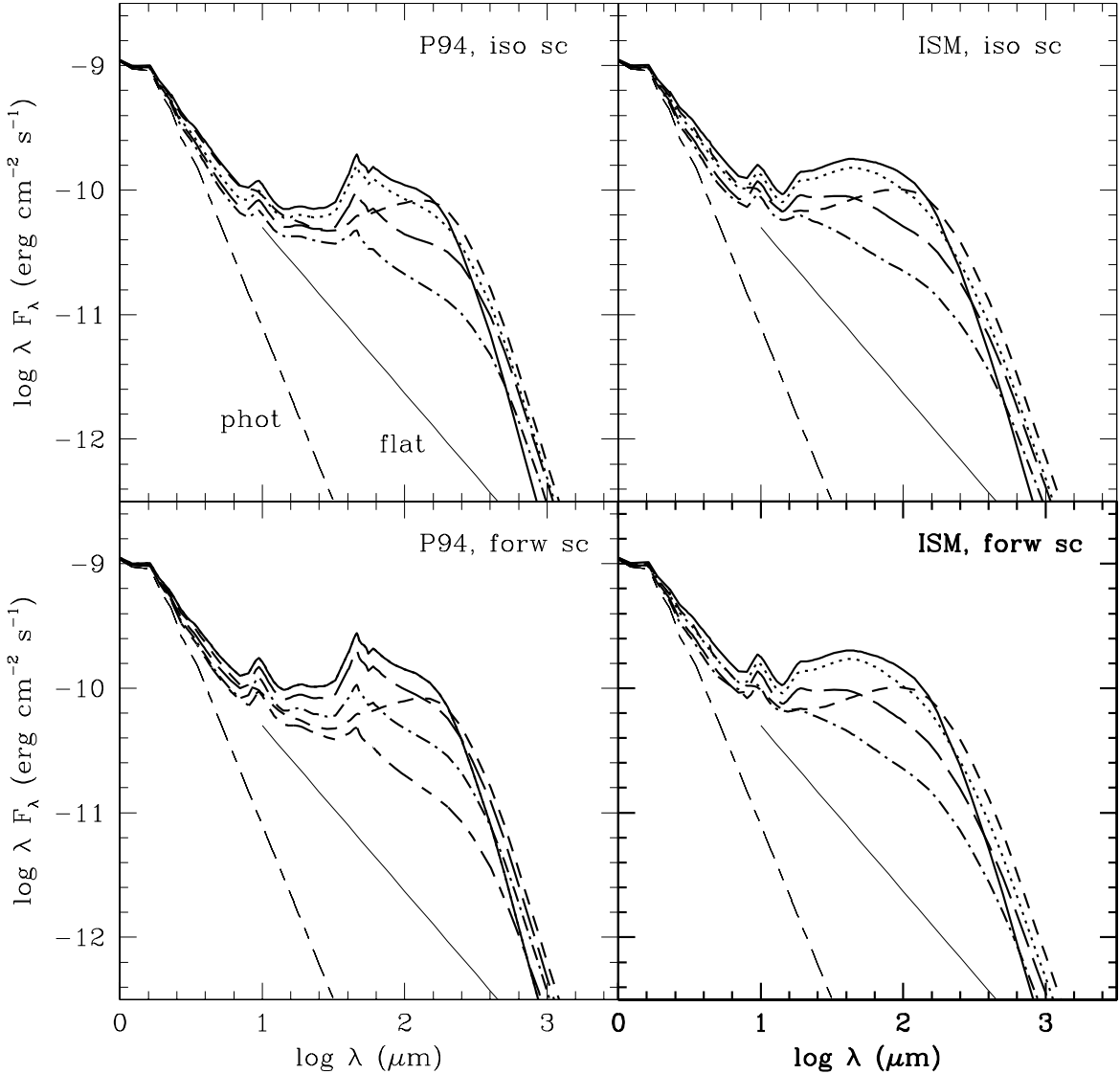}
\caption{SEDs of models with 
ISM dust (right panels) and 
P94 dust (left panels),  
and different depletions $\epsilon$ = 1 (solid line),  0.1 (dotted line),  
0.01 (dot-dashed line),  0.001 (long-dashed line).
For comparison, 
the SEDs of a well mixed models with 
%$a_{max}=0.25 \ \mu m$ (dotted line) and 
$a_{max}=$ 1 mm (short-dashed line) is shown. 
The SED for a well mixed model with $a_{max}=0.25 \ \mu m$
is similar to that of the model with $\epsilon= 1$.
SEDs in upper panels are calculated assuming 
isotropic scattering of the stellar 
radiation and SEDs of lower panels, assuming perfectly forward scattering.
The photosphere is shown in light short-long dashed lines, and 
a light solid line with a slope -4/3, corresponding to the
spectral slope of a flat disk,
is shown as reference.
}
\label{fig_sed}
\end{figure}

Figure \ref{fig_sed} shows 
the SEDs for models calculated with the fiducial parameters
and with the two assumed dust
compositions. Models are calculated for four values of
the depletion parameter $\epsilon$ and the two limiting cases of
isotropic and completely forward scattering of stellar radiation
(\S \ref{secc_models}). 
The SED of the central star was taken from Bruzual \& Charlot (1993).

For both dust mixtures and assumptions of scattering properties,
the SEDs show a similar general behavior,
also described in Dullemond and Dominik (2004). As $\epsilon$ decreases,
the emission at infrared wavelengths decreases, and the 
infrared SED becomes stepper.
As discussed, this is a consequence of how the irradiation surface changes when
the atmospheric dust is depleted (see Figure \ref{fig_superf}).
A lower $z_s$ implies a lower irradiation flux entering
the disk and thus a lower emergent flux.
However, even for the case of the highest depletion we are considering,
the spectral
index in the 25 $\micron$ - 100 $\micron$ region is flatter
than that of a perfectly flat disk, $n \sim  - 4/3=-1.33$ (see Fig. \ref{fig_sed}). 
In other words, the disk is flared even for a dust
to gas mass ratio of only 0.1 \% the standard value, 
The slope of the SED in the mid-IR and far-IR depends on the
dust composition in addition to $\epsilon$. 
For instance, the water ice in the P94 mixture produces in
emission bands at $\sim$ 50 - 60 $\micron$ (cf. Fig. \ref{fig_opas}), which tend
to flatten the SED in this wavelength region relative to the disks with ISM dust.

Figure \ref{fig_sed} also shows
that the mm flux first increases between $\epsilon$ = 1 and 0.1,
and then decreases with $\epsilon$.
Since the opacity at mm wavelengths is low,
the largest contribution to the flux in the mm comes from 
the dense dust layer near the midplane.
Although the dust to gas mass ratio of the big grains layer
increases when $\epsilon$ decreases (Table \ref{tabla_epsilon} ), 
the lower temperature
at the midplane (Figure \ref{fig_temperaturas}) results in lower mm fluxes.  
The mm slope of the SED becomes flatter
because the temperature drops to low enough values that the Planck
function does not have precisely the Rayleigh-Jeans slope.

Even models with very flat SEDs
show emission in the 10 $\micron$ silicate feature
(Figure \ref{fig_sed}). 
The ratio between the flux at the center
of the feature and to the interpolated
continuum flux depends on
the type of silicate, the assumed dust composition,
the asymmetry factor of the grains, and the
degree of settling, in addition to the inclination to the
line of sight.
ISM dust produces a stronger band than P94 dust if scattering is
isotropic. The main reason is that the upper atmosphere of the disk
with ISM dust is hotter than the case with P94 dust (see
Figures \ref{fig_temperaturas} and \ref{fig_vertical}).
On the other hand, P94 dust has a higher albedo than ISM dust,
so the silicate feature flux increases more for P94 than for
ISM dust as scattering becomes less isotropic.

\begin{figure}
\plotone{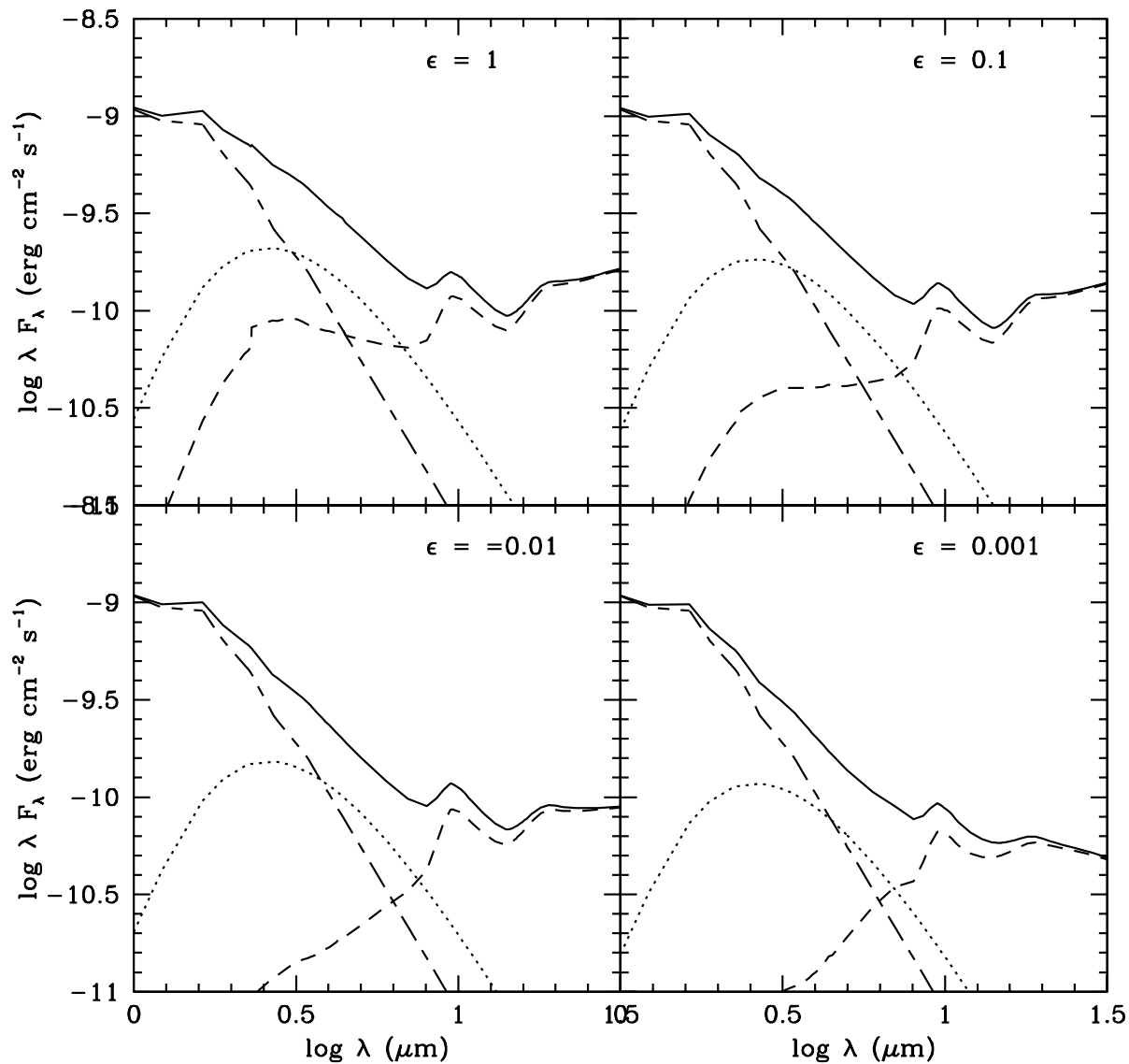}
\caption{SEDs of  the fiducial model with ISM dust, isotropic scattering, 
and four values of $\epsilon$. The contributions
of the disk (dashed line), wall at the dust destruction radius
(dotted line), and photosphere (short-long dashed line) are shown
separately. 
}
\label{fig_flux}
\end{figure}

In Figure \ref{fig_flux} 
we show the SEDs of Figure \ref{fig_sed} with ISM dust
in the 1 - 40 $\mu$m
region, indicating the the separate contributions of the wall
at the dust sublimation radius, the disk truncated at this
radius, and the stellar photosphere. The wall is a major
contributor to the flux shortward of $\sim 10 \mu$m. At the inclination 
of the fiducial model ($\mu$ = 0.65,
$i = 50^o$) and for this type of dust, the contributions of the 
wall and the disk are similar
at $\lambda \sim 6-8 \mu$m, and the wall dominates the excess above the
photosphere at shorter wavelengths. The wall contributes $\sim$ 20 \%
of the flux at 10 $\mu$m. As $\epsilon$ decreases, the excess above
the photosphere decreases.

\subsection{Spatial distribution of the emergent flux}
\label{secc_spatial}

\begin{figure}
\plotone{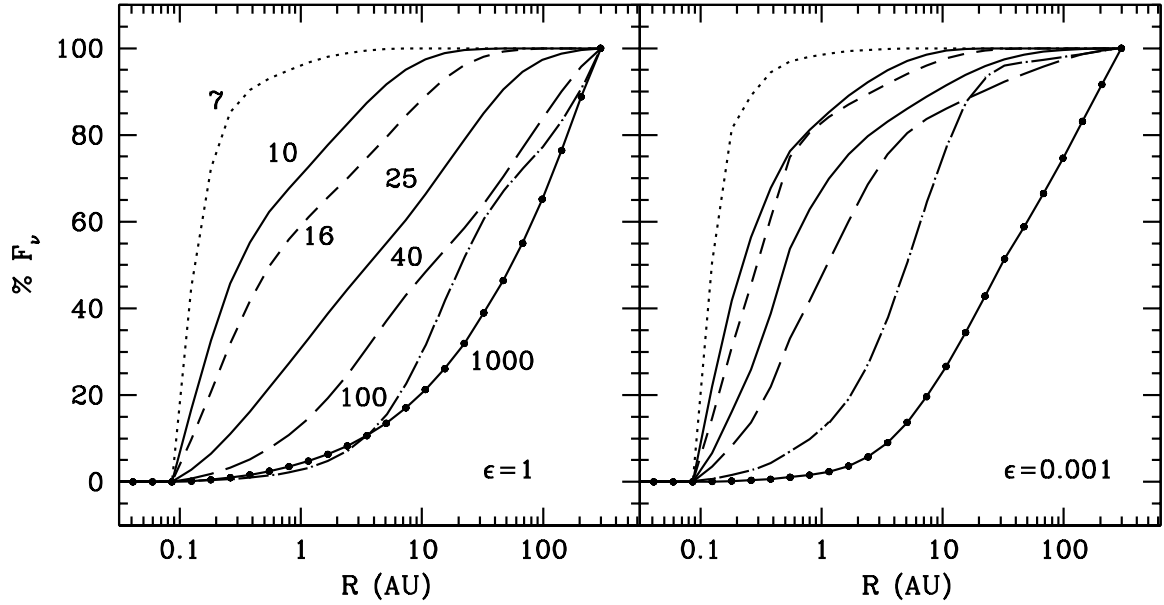}
\caption{
Spatial distribution of the emergent 
flux from the disk 
at different wavelengths (each curve is labeled with the wavelength in 
$\mu$m), for the fiducial model with
%two disk models: 
$\epsilon=1$ (left panel) and $\epsilon=0.001$ (right panel).
The models have ISM dust and isotropic scattering
and the disk radius is 100 AU.}
\label{fig_spatial}
\end{figure}

Figure \ref{fig_spatial} shows 
relative cumulative flux at radius $R$ as a function of radius
at representative wavelengths 
for the  fiducial model with ISM dust and
two degrees of settling,
$\epsilon=1$ and $\epsilon=0.001$, and pole-on disks
($\cos i=1$.)
In general, outer regions contribute more flux at longer
wavelengths, but the actual contribution of a given region to the
flux at a given wavelength depends on $\epsilon$.
For both values of $\epsilon$, the short wavelength side 
of the 10 $\mu$m band, emerges from  $R < 0.2$ AU,
but for longer wavelengths the contribution of regions outside 10 AU becomes
less important as $\epsilon$ decreases.
For $\epsilon = 0.001$, a fraction $\ge$ 80 \% of the
flux up to 100 $\mu$m arises from $R \leq 10$ AU,
and $\sim$ 50 \% of the millimeter flux is formed
inside 30 AU. 

In addition to continuum points, Figure \ref{fig_spatial} shows
the region of formation of the 10 $\mu$m flux, due 
to silicate emission.
Again, the extent of this region depends on $\epsilon$.
For $\epsilon$ = 1, 70 \% of the flux at 10 $\micron$ at forms inside 1 AU,
while for $\epsilon$ = 0.001, 85 \% forms in this region;
in both cases all the flux at 10 $\micron$ arises at $R < 10$ AU.

The observed flux at 10 $\micron$ will be a sum of emission from the disk
and the wall at the dust destruction radius. For
$\cos i=1$ there is no wall contribution because it is assumed to be
vertical, but for $i = 50^o$, the wall
contribution is $\sim$ 20 \% (\S \ref{secc_emission}),
and it becomes larger as inclination increases 
(until the wall begins to self-shadow, $i \sim 70-80^o$).

\subsection{Radial dependence of $\epsilon$ }
\label{secc_veps}

\begin{figure}
\plotone{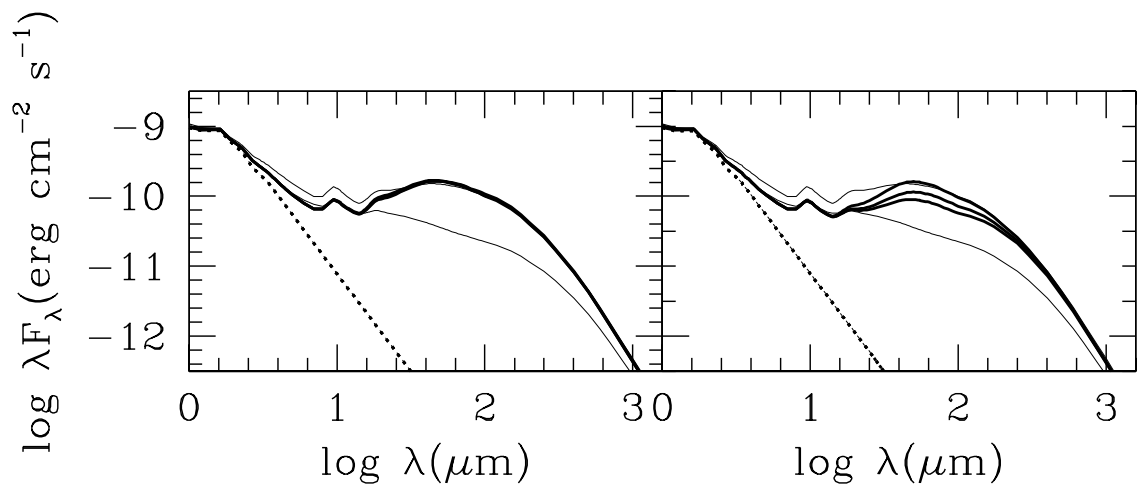}
\caption{SEDs of models with $\epsilon(R)$. 
Models with $\epsilon_1$ = 0.01 and $\epsilon_2$ = 0.1,
spanning different regions in the disk.
Left: $R_1$ = 10 AU and $R_2$ = 50, 75, and 90 AU, thick
lines from top to bottom.
Right: $R_1$ = 50 AU and $R_2$ = 75, 100, and 200 AU, thick
lines from top to bottom.
For comparison, 
in the two panels we show SEDs of models with constant $\epsilon=$ 0.1 and 
0.01 (upper and lower light solid lines).
}
\label{fig_veps}
\end{figure}

In this section we study the effect of changing the depletion of
the upper layers with radius. 
To restrict the parameter space, we adopt the following functional form,

\be
\epsilon(R)=
\cases{ 
               \epsilon_1 &  if $R < R_1$, \cr
               \epsilon_1 (R/R_1)^f & if $R_1 < R < R_2$, \cr
               \epsilon_2 & if $R > R_2$
}
\en

where the exponent of the power law is 
 $f=[\log(\epsilon_2)-\log(\epsilon_1)]/[\log(R_2)-\log(R_1)]$.
We adopt $R_1 < R_2$ and
 $\epsilon_1 < \epsilon_2$, because we assume 
that dust settling in the inner disk is faster 
 than in the outer disk, since the timescale for settling in a quiescent disk 
is comparable to the orbital timescale (Weidenschilling et al. 1997;
Dullemond \& Dominik 2004).
Different combinations of the four parameters $\epsilon_1$, $\epsilon_2$, 
$R_1$ and $R_2$ produce different SEDs and 
Figure \ref{fig_veps} shows some examples. 

In Figure \ref{fig_veps}, we keep 
$\epsilon_1$ and $\epsilon_2$ fixed at 0.01 and 0.1, respectively,
and
show the effect of changing the
extent in the disk of the regions with different settling.
For comparison, we show models with the corresponding constant
values of $\epsilon$. In the left hand side,
the inner most settled region extends to 10 AU, and
the degree of settling increases to radii 50, 75, and 90
AU. It can be seen that the fluxes up to $\sim$ 15 $\mu$m
correspond to those of the model with $\epsilon$ = 0.001,
while for $\lambda > 25 \mu$m, the fluxes correspond
to those of the model with $\epsilon$ = 0.1. This is in 
agreement with the fact that for the large values of
$\epsilon$ of the outer regions, most of the flux at 
$\lambda > 25 \mu$m comes from regions around 100 AU
(see Fig. \ref{fig_spatial}). In the right hand side.
the more settled inner disk regions extend to 50 AU,
while the transition region extends to 75, 100, and 200 AU.
In this case, there is a larger effect on the far-IR
fluxes, because the outer disk regions where most of this
flux is produced are affected.

\subsection{Effect of the grain size distribution in the upper layers}
\label{secc_amax}

So far, we have shown models in which the upper layers have
 a fixed maximum grain size of $a_{max}=0.25 \ \mu m$.
This value was chosen because it is characteristic for the diffuse ISM
(e.g., DL). However, grains might have grown
in the cores where disks are formed (e.g. Weidenschilling
\& Ruzmaikina 1994) and/or small grains might have been blown away
from the disk upper layers by radiation pressure.
On the other hand, grains with radius $a=0.25 \mu m$ and smaller
might have already settled, leaving only the smallest grains
at the upper layers.
Figure \ref{fig_sed_amax} shows the effect on the SED of
values of $a_{max}$ = 0.05, 1, and 10 $\mu$m in the upper layers, 
for $\epsilon=$ 0.1 and 0.001.
We can see that the smaller the grains the stronger the bands,
because the grains in
the upper atmosphere are hotter when $a_{max}$ decreases.
The $10 \ \mu m$ silicate band almost disappears
for $a_{max} > 1 \ \mu m$.
Also, for a given $\epsilon$,
smaller grains result in higher IR continuum, reflecting
the fact that the small grains absorb the stellar radiation
much more efficiently.

\subsection{Effect of the height of the layer with large grains}
\label{secc_zbig}

We have assumed so far that 
the layer of big grains is geometrically thin.
This situation might be expected if the disk is old and quiescent enough,
such that settling has taken place in the whole disk vertical
structure fairly homogeneously.
However, during the first stages of settling, only grains
from the regions with the lowest density may have settled.
Also, as shown by Miyake \& Nakagawa (1995), and more recently by
Dullemond \& Dominik (2004), the turbulence described by an $\alpha$
prescription may be  able to  mix grains
up to several scale heights.

In this section we explore the effect in the SED of adopting different
heights for the big grains layer, $z_{big}$ (see Figure \ref{fig_geom}).
We parameterize the
height of the big grains layer as a constant times the local gas
scale height $H=c_s(T)/\Omega_k$.
The assumption that the layer of big grains at the disk midplane
is  geometrically thin is relaxed, and
we calculate $\zeta_{big}$ for each $z_{big}$  consistently
for each value of $\epsilon$ (see Appendix).
As described in papers I, II and II, the calculation of the
disk structure is performed using an iterative scheme
in order to obtain the irradiation surface
and the disk structure self-consistently.
In the case of $z_{big}/H $ of order 1 or larger
we numerically integrate the density over the different disk regions 
in each one of these iterative steps. 

Figure \ref{fig_sed_zbig} shows SEDs for the fiducial model with ISM dust,
isotropic scattering, $\epsilon=0.1$ and 0.01, and different values
of $z_{big}$.
The upper left panel shows the cases with $z_{big}=0.1 H$, for
comparison. As shown in the upper right panel, SEDs for $z_{big} \le 1 H$
are very similar to the thin layer case, because the irradiation
surface is still well above $z_{big}$,  and its location 
depends on the small grains and their depletion; in these cases,
the optically thin emission from the upper hot atmosphere is
dominated by the emissivity of the small grains.
However, as $z_{big}$ increases and $\epsilon$ decreases,
the irradiation surface moves closer to the region with large
grains. 
The near-IR SED of cases with $z_{big}=2 H$ corresponds
to that of an atmosphere with small grains, because
at small radii the
irradiation surface is still above $z_{big}$; however,
the far IR SEDs tends toward the SED of the well mixed disk with $a_{max}=$ 1 mm,
becoming more similar as $\epsilon$ decreases.
In the extreme case of $z_{big}=3 H$, the two models
with different values of $\epsilon$ have the same SED than the
well mixed model with $a_{max}=$ 1 mm and no depletion.
In this case
the dust mass of the layers above $z_{big}$ is negligible compared
to the dust mass below, and  the dust to gas mass ratio of the
big grains layer is close the standard value.

\subsection{Dependence on mass accretion rate}
\label{secc_mdots}

Our models are for
{\it accretion disks} irradiated by the central star.
The fact that these are accretion disks enters in three ways into 
our model calculations. First,
viscous dissipation is considered an additional disk heating source.
This heating mechanism is important close to the star and close to the midplane, 
to an extent dependent upon the ratio of $L_{acc}$ to $L_*$.
In those regions where  viscous dissipation significantly alters the midplane 
temperature, the gas scale height also changes, affecting the 
whole vertical density distribution and, consequently, the height
of the irradiation surface and the reprocessed stellar flux.
In addition, the energy flux released by viscous dissipation modifies
the SED with respect to a purely reprocessing disk.
Second, the inner edge of the dusty disk is irradiated both by the star
and by the accretion shocks at the stellar surface, where disk material is 
channeled onto the star by the stellar magnetosphere (Muzerolle et al. 2003; 
D'Alessio et al. 2003).
Third, the disk density depends on the mass 
accretion rate through the conservation of angular momentum flux. 
The disk mass surface density density is proportional to 
$\Mdot/\nu_t$, where $\nu_t$ is a density weighted mean
 turbulent viscosity coefficient (c.f., \S 3.1).
Thus, the higher the mass accretion rate, the higher the disk density, and the 
higher the irradiation surface.
As a consequence, disks with higher mass accretion rates will have 
higher fluxes,  even if turbulent viscous dissipation is not as important as
stellar irradiation as the disk heating mechanism.
Correspondingly, disks of the same outer radius and the same $\alpha$
with higher accretion rates will have larger masses. 

Figure \ref{fig_mdots} shows the effect on the SED of changing the 
mass accretion
rate and the  degree of settling. It clearly can be seen that
the mid and far IR  and 
sub-mm and mm fluxes scale with $\Mdot$  for a given value of $\epsilon$.
It can also be seen that even the case with the largest mass accretion rate shown, 
the emergent SED depends on $\epsilon$, because irradiation
is still an important heating mechanism.
As previously shown by Calvet et al. (1991), the intensity of the 10 $\mu$m 
silicate band decreases as mass accretion rate increases, because of the 
decrease 
in the contrast between the temperatures at the silicate band formation height
and the continuum formation height. However, the present models do not include
the disk irradiation by the accretion shocks at the stellar surface, which 
might be an important irradiation source for $\Mdot \gtrsim 10^{-7} \ \MSUNYR$ 
(D'Alessio et al. 2004) and 
which 
probably would increase the gradient of temperature in the disk atmosphere.
The role of this irradiation source will be discussed in a forthcoming paper.

\subsection{Dependence on inclination angle}
\label{secc_angles}

Figure \ref{fig_angles} shows the SEDs of 
the fiducial model with ISM dust
for 4 different high inclination angles: 
$\cos i =$ 0.1, 0.25, 0.35 and 0.5 
(i.e., $i=$ 84, 75.5, 69.5 and 60 $^\circ$), and different 
degrees of settling: $\epsilon=$1, 0.1, 0.01 and 0.001.
All the SEDs include the contribution of the star and the wall, 
the thermal emission from the disk, and the contribution
of the stellar light scattered in the disk atmosphere and the disk 
thermal emission. 

At high inclinations, the flared outer disk regions
absorb radiation 
from the star, wall, and inner disk. 
For a given inclination,
disk surface density distribution and outer radius
the resultant extinction
depends on $\epsilon$; as the disk becomes more settled
and geometrically flatter, the range of inclination
angles for which the outer disk occults the star is narrower.

In the models with the highest inclination angle in the
upper right corner of Figure \ref{fig_angles},
$\cos i =$ 0.1, 
the star contributes only indirectly to the SED through scattered light
in the case $\epsilon$ = 1.
This produces a
characteristic double-bump shape. Notice that the thermal emission of
an highly inclined disk might be similar to that of a class I/0 source
(Chiang \& Goldreich 1999) but the contribution of the scattered light
makes the total SED very different to a class I/0 source and even to
a class II source (see paper II).
As $\epsilon$ decreases and the outer disk
becomes flatter, 
the peak of the thermal emission moves toward
shorter wavelengths and the silicate absorption feature becomes shallower,
reflecting the decreased opacity of the outer regions. At the same time, 
the lower the $\epsilon$, the lower the far-IR
flux of  the disk as it intercepts less radiation from the star.
All these edge-on models have similar contributions of scattered
stellar light, since
it depends on the albedo which is independent on dust to gas mass ratio.

The rest of the panels in Figure \ref{fig_angles} show the SEDs
for lower inclinations. Occultation effects including silicate absorption
are still present,
but they only affect the less settled objects as inclination decreases.
Mid- and far-IR continuum fluxes increases roughly proportional
to $\cos i$, as expected for an optically thick, geometrically flat disk,
while the optically thin millimeter fluxes are independent of inclination.

\section{Observational tests}
\label{sec_obs_tests}

\subsection{Median SED of CTTS in Taurus}
\label{secc_median}

Figure \ref{fig_median} compares the predictions of our fiducial settled models
with median fluxes for the Taurus population. For $\lambda < 10 \mu$m,
the median values and quartiles were constructed from the 2MASS and
IRAC magnitudes of Hartmann et al. (2005). This sample consists of
19 stars.  Fluxes were scaled to the H band,
after correcting for reddening as in Furlan et al. (2005b). At longer
wavelengths, we show the median from Paper II.
We also show in Figure \ref{fig_median} predictions for 
the fiducial model with ISM dust 
and two values of $\epsilon$ = 0.1 and
0.01. The models are shown for two inclinations, 30$^\circ$ and 60$^\circ$ ($\mu$ = 0.87 and 0.5),
because as already mentioned, the median may corresponds to inclinations
lower than $\mu$ = 0.5 if the highly inclined objects are too extincted
to be included in the sample. The models have been corrected for self-extinction,
to make a consistent comparison with a sample corrected for reddening.

For $\lambda < 6 \mu$m, model predictions are roughly consistent with
the IRAC median.  The models are slightly below the median values
at the IRAC bands, even though
we have assumed no settling in the inner wall 
to increase the wall emission and thus short-wavelength fluxes.
Several factors may play a role in accounting for these differences.
First, photospheric fluxes have a large contribution at these wavelengths,
and details of the adopted SEDs can affect the comparison. 
Here we have added the excess disk + wall fluxes to the 
dereddened fluxes of the Weak (non-accreting)
T Tauri star HBC 274 from Hartmann et al. (2005), which may be
more representative of the active photospheres of the stars.
Second, if true accretion rates are higher than the
assumed $\mdot = 10^{-8} \msunyr$, the near-infrared fluxes would
increase.  As can be seen in Figure \ref{fig_mdots}, 
the near-IR flux increases more rapidly for $\mdot > 10^{-8} \msunyr$
than it decreases for $\mdot < 10^{-8} \msunyr$. This is due
to the fact that for $\mdot < 10^{-8} \msunyr$, the radius of the
wall is effectively set by the stellar luminosity, while 
for larger rates, the wall radius increases as
the accretion luminosity increases and becomes more comparable
to the stellar luminosity. 
The uncertainty in the mass accretion rate determinations
is probably a factor 2 or even perhaps 3; increasing the mean
accretion rate would improve the agreement with the median SED.
Third, the geometry of the wall may play a role.
We have assumed that the wall is vertical for simplicity, but
the actual geometry may be
more rounded because of the height dependence of the density 
(Isella \& Natta 2005).
If this is this case, the more area of the wall may be exposed and
the inclination dependence we have used is no longer valid.
A more detailed study of the inner disk regions, including
information from interferometric measurements (Akeson et al. 2005; Eisner et al. 2005),
may clarify these points further.

Although the models appear to be low at $8 \mu$m, this
cannot be trusted, because the IRAC band is not a monochromatic
measurement but instead includes some silicate emission
in its bandpass.  Detailed comparisons require the data
forthcoming from the Infrared Spectrograph on the Spitzer
Space Telescope.
In a separate paper (D'Alessio et al. 2005c, in preparation) we
carry out a detailed exploration of the effects of different
compositions of the dust, in particular of the silicates, and
show that the silicate feature can be reproduced in individual
cases assuming a large variety of compositions, in agreement
with detailed determinations by B. Sargent et al. (2005, in preparation).

The models reproduce in general terms the long wavelength median SED,
as already pointed out by Furlan et al. (2005a). 
While it appears that the mid-IR is better reproduced by 
models with $\epsilon=0.1$ with smaller values of $\epsilon$ are 
required at longer wavelengths, it should again be noted that
the median SED in this region is derived from IRAS observations,
and the comparisons will need to be revisited once detailed
{\em Spitzer} results become available.

In addition, the model fluxes
in the millimeter range are slightly lower than the median
fluxes. This may be remedied by models where
the layer with big grains covers a larger height in the disk. 
to illustrate this, we show in 
Figure \ref{fig_median} a model with $\epsilon=0.1$ and
$z_{big}$ = 2 H from Figure \ref{fig_sed_zbig}, which seems to improve
the comaprison.
In any event, more information about degrees of settling in real objects
require detailed modeling of individual SEDs.

\subsection{IRAC colors of Taurus CTTS}
\label{secc_irac}

Figure \ref{fig_irac} shows IRAC colors of CTTS in Taurus from
Hartmann et al. (2005). The central wavelengths of these
bands are 
3.6 $\mu$m, 4.5 $\mu$m, 5.8 $\mu$m, and 8 $\mu$m,
covering the wavelength region
shortwards of the silicate feature with an important
or dominant contribution from the wall at the dust destruction
radius (except at extremely low or high inclinations
to the line of sight). 
As discussed, this wavelength region is actually probing
the innermost disk
(Figure \ref{fig_spatial}).  
Model colors have been constructed
by adding excess emission fluxes (disk + wall) integrated over
the band transmissions to those of the Weak T Tauri star HBC 274 from 
Hartmann et al. (2005), of
spectral type K7 scaling to the same 
luminosity as the model, 
to make sure that the colors tend to observed 
photospheric colors when the excess tends to zero.

Models are shown for several values of mass accretion
rate, settling parameter, and inclination (see caption
to Figure \ref{fig_irac}).
For a given value of $\epsilon$, predicted colors
increase with mass accretion rate, as the emission from the
disk and specially the wall at the dust destruction radius
increases (\S \ref{secc_emission}, \S \ref{secc_mdots},
and Figure \ref{fig_mdots}).

The largest inclination dependence of colors
is seen for models with high mass accretion rate
and/or large values of $\epsilon$. This  is due
to the effects
of disk self-absorption in these optically
thicker models (\S \ref{secc_angles} and Fig. \ref{fig_angles}).
In particular, models at the higher inclination shown in Figure \ref{fig_irac}
(cos $i$ = 0.25)
tend to have
very {\it red} [3.6]-[4.5] color and
very {\it blue} [5.8]-[8] color; the SED of 
the $\epsilon$ = 1 model in upper right
panel of Fig. \ref{fig_angles} has these
colors: a very red slope shortwards of the silicate
feature, and a deep silicate absorption, which
results in a low flux in the broad IRAC [8],
and in a blue [5.8]-[8] color.

Colors [3.6]-[4.5] and [4.5]-5.8] 
become bluer as the settling parameter decreases,
reflecting the decrease of flux of the
the inner disk. 
The reddest values of the [5.8]-[8] color are not
predicted by the models. Since the [8] band has a large
contribution from the silicate emission feature, this
is the same problem encountered when discussing the median
SED (\S \ref{secc_median}).

Overall, the observed range of IRAC colors in 
Taurus can be explained by the predictions of settled models;
settling has only modest effects on the colors in this wavelength
range.
Slightly better agreement is obtained at somewhat higher
mass accretion rates than the fiducial $\mdot = 10^{-8} \msunyr $.
It is difficult to compare our results with those of
Whitney et al. (2003, 2004) because they exhibit very few disk models for
low-mass stars. 

\subsection{Mid-IR spectral slopes}

Furlan et al. (2005a) calculated slopes of the fluxes
in {\em Spitzer} Infrared Spectrograph (IRS) spectra of a sample 
of CTTS in Taurus, using continuum bands (i.e., free from features)
 of spectra from Spitzer IRS. The center wavelength of the selected bands
are 5.7, 7.1, 13.25, 16.25 and 25 $\mu$m. 
Figure \ref{fig_irs} shows the logarithm of the ratio of fluxes
13/25 and 6/25 (rounding up with central wavelengths
of the bands to the nearest digit), denoted by indices,
from Furlan et al. (2005a) compared to predictions for the fiducial 
model with ISM dust with different mass accretion rates,
inclination angles, and values of $\epsilon$.

Except for models with $\epsilon = 1$,
the value of the indices increases with inclination. 
This
is due to the fact that as inclination increases the wall emission increases
(up to $\sim 70^\circ$), increasing the 6$\mu$m flux and to a
lesser extend the 13$\mu$m flux, while disk emission at 25$\mu$m decreases
(Fig. \ref{fig_angles}).
For models with $\epsilon \sim $ 1, the short wavelengths
are self-absorbed and the indices decrease.  

Most of the observations fall in the region corresponding
to mass accretion rates $10^{-9} \ \MSUNYR$ and $10^{-8} \ \MSUNYR$,
and inclinations $30^\circ$ and $60^\circ$, as expected.
Moreover, as already discussed in Furlan et al. (2005a)
the observations are consistent with values of settling
$\epsilon \le 0.1$. 

\subsection{Scattered light images of edge-on disks}
\label{secc_edge}

In Paper III we showed that disk models with larger grains have less 
opacity to the stellar radiation producing flatter images in scattered 
stellar light when disks are seen edge-on. 
The flattening of the near-IR  images of edge-on disks with respect to what is 
predicted if disks have ISM-like dust (c.f., paper II) was required to 
explain cases like HH 30 and HK Tau B.

As expected, a similar flattening of the image results 
from increasing the depletion of small grains in the upper layers.
Figure \ref{fig_scatt} shows how the distance between the bright  
nebulae  and the width of the dark lane decrease when $\epsilon$ 
decreases, because the upper layers becomes more transparent to 
stellar radiation.
We do not attempt a detailed modeling of any object at this time, but 
aim to show that the image of an 
edge-on disk  with certain degree of settling is consistent 
with  the image of an edge-on disk with well mixed big grains.
It is possible to distinguish between both cases when there is 
information available at several wavelengths. For instance, combining
scattered light images, a complete SED, and 
detailed modeling, it seems possible to quantify the upper and 
lower grain sizes and the degree of settling of a given object, 
if its mass accretion rate and central star properties are known.

\section{Discussion}
\label{sec_dis}

\subsection{Dependence on assumptions}

Within the context of our assumptions of an initial
(protostellar) dust distribution consistent with that of
the diffuse ISM, power-law grain size distributions, and
no radial dependence of dust properties, we can
only explain the combination of high mm-wavelength emission
and ubiquitous $10 \mu$m silicate emission with a combination
of both dust growth (to explain the long-wavelength fluxes)
and settling (to move large grains to the midplane, leaving
behind small grains to produce strong silicate features),
as suggested in Paper III.  In principle, we could also
obtain this result without settling
by having the large grains only at large disk radii, so
they can dominate the long-wavelength emission but not
contribute to the silicate features; however, this seems
unlikely given that dust evolution is likely to be fastest
in the inner disk.

Quantifying the degree of settling in a disk 
from observations is much more difficult.
Models of dust evolution suggests that the timescale for
settling is proportional to the orbital time (Weidenschilling 1997;
Dullemond \& Dominik 2004); however, these models predict a much
rapid evolution of dust sizes and faster settling than consistent 
with what is observed in systems of ages of a few Myr.
Other factors such as turbulence,
grain shape, and a more complete treatment of particle interaction
may be required to obtain closer agreement with observations
(Dullemond \& Dominik 2005).
Given all the theoretical uncertainties, the parameterized treatment
given here can provide a useful comparison with observations.

In \S \ref{secc_overview}, we discussed that the degree of depletion
in the disk upper layers
required to fit the slope of the SED - indicative of large grains - and at the same time have 
silicate emission - requiring small grains - was set by 
the condition that the opacity in the wavelength range where the stellar
radiation is absorbed, due to the small grains in the upper layers,
had a similar value to that of large grains.
This condition can be attained either by
(1) decreasing the dust to gas
mass ratio in the upper layers or 
(2) decreasing the opacity per 
small grain at the stellar wavelengths,
while keeping the opacity at longer wavelengths similar,
because for a given dust to gas mass ratio different combinations of  dust ingredients,
optical constants, grain sizes and shapes can produce  different
wavelength dependences of the opacity.
Can we distinguish between these two possibilities?

The continuum flux 
emerging from the disk interior regions, for a given central star, 
disk inclination angle, and mass accretion rate, depends on  
the angle between the incoming radiation and the normal to the
disk surface, which is determined by the mean opacity to the stellar radiation. 
In both possibilities above, 
depletion or decreased opacity per 
dust mass,
the disk would have the same irradiation surface, and thus 
similar flux from the interior layers.
However, the flux emerging from the optically thin upper hot layers
would be different.  In the case of 
lower opacity per grain, the temperature of the upper layers
$T_0$, given by eq.(\ref{tthin}) would be lower, since it depends
on the ratio $(\kappa_*/\kappa_P)^{1/4}$,
and $\kappa_P$ is 
nearly constant for $T < 1000K$ (D'Alessio 1996). 
The emission of these layers,
as given by eq.(\ref{ithin}) will also decrease although it
depends on $1/\chi_*$, because the decrease due to the exponential dependence
on temperature in $B_\nu(T_0)$ would dominate.
Since the emission from the upper layers constitutes
a significant fraction of the total flux in the mid-IR (Chiang \& Goldreich 1997),
the case with high $k_*$ and depletion has a larger mid-infrared flux
than the case of lower opacity per 
dust mass, essentially because the small dust is hotter. 
This means that in principle, the two possibilities could be distinguished
by matching the overall SED of actual objects.

A further complication is related to the 
way in which the optically thin dust temperature is calculated. 
It is now accepted that in the uppermost layers of the disk, 
where densities are very small, 
the gas temperatures is higher than the dust temperature (Glassgold et al. 2004).
Moreover, different dust ingredients and grains sizes are characterized 
by  different temperatures
since each grain is heated mostly by the absorption of the stellar radiation 
it intercepts, which depends on its specific absorption coefficient (e.g., Li \& Greenberg 1998). 
However, close to the irradiation surface the disk density 
is high enough to assure thermal coupling between gas and dust, and that the
main heating mechanism is the dust absorption of stellar radiation
(Kamp \& Dullemond 2004). 

\subsection{Impact of settling on disk ionization structure}

Sano (1999) and Sano et al. (2000) considered the 
effect of the dust in calculating the 
ionization state of protoplanetary disks,
with the goal of understanding the location and size
of the region 
where the magnetorotational instability can operate. They find 
that with a interstellar values for the dust to gas mass ratio and 
the grain sizes, and assuming a standard cosmic ray
ionizing flux, the ionization fraction is such that Ohmic dissipation 
makes the disk stable inside $\sim $ 20 AU. However, 
they also find that dust growth and depletion decreases the size 
of this region, allowing turbulence to be 
generated by the magnetorotational instability in most of the disk.
This occurs because the small dust particles are particularly
effective in enhancing the recombination of electrons and ions.
In all their models there is a remaining {\it dead zone} (Gammie 1996) 
close to the star, with a size that depends not only on $\epsilon$ and $a_{max}$ 
but also on the disk mass surface density.  The fiducial disk model
of Sano et al. (2000) has the 
minimum solar mass surface density, which 
has a steeper dependence on radius than in our models,
and thus is higher than 
the $\alpha$ accretion disk mass surface density for $R < 100$ AU
(if $\Mdot=10^{-8} \ \MSUNYR$ and $\alpha=0.01$). 
Sano et al. show that lowering the mass surface 
density also has the effect of decreasing the size of the dead zone. 
A rough comparison with Sano et al. results suggests that 
a model with $\Mdot=10^{-8} \ \MSUNYR$, $\alpha=0.01$ and $\epsilon=0.01$  
would have a dead zone a factor of 0.1 smaller than their 
fiducial model 
(i.e, at $R < 2$~AU).  A detailed calculation of 
the ionization state of a settled accretion disk model is required
to quantify the limits of the stable region properly.

\subsection{Dust settling and shadowed disks}

The possibility of disks being self-shadowed has been raised
in a series of studies of disks around Herbig Ae/Be stars (Dullemond \& Dominik 2004a,
and references therein), and more recently in the
exploration of dust settling by Dullemond \& Dominik (2004b).
This last study was the first to point out the
main effect of dust settling on the SED of protoplanetary disks,
namely, the decrease of mid to far-IR excess with dust settling.
One of the inferences of the Dullemond \& Dominik (2004b)
study is that the outer portions of the very settled disks are self-shadowed;
similarly, we find that the disks with the larger degree of
settling in our study, $\epsilon <$  0.01, become optically
thin to the stellar radiation outside $\sim$ 100 AU (Fig. \ref{fig_superf}),
so that the disk surface  is no longer flared in those outer
regions. However, one important point to notice is that
the decrease of IR fluxes is not due to
the fact that those outer regions are shadowed by the inner
disk regions. As discussed in \S \ref{secc_spatial} and shown in
Figure \ref{fig_spatial},
as $\epsilon$ decreases,
an increasing contribution of the flux comes from
the innermost regions of the disk, which are optically
thick to the stellar radiation and so still have a flared surface.

Shadowing effects are more likely to be present
if the surface
density decreases more steeply with radius than $\sim R^{-1.5}$, as
shown for the case of the Herbig Ae/Be stars by
Dullemond \& Dominik (2004a). If this is the case,
the density in the outer disk becomes low enough to render
it optically thin to the stellar radiation.
However, in our studies,
the surface density distribution is not a free function,
but is it calculated self-consistently from the
mass accretion rate, the $\alpha$ parameter, and the disk temperature
distribution. From these calculations, we obtain not
only the absolute value of $\Sigma$, consistent
with the mass accretion rate, but also the dependence
on radius. From Fig. \ref{fig_sigma} we see that
$\Sigma \propto R^{-1}$ for most of the disk;
even for the most settled disks, $\Sigma \propto R^{-1.2}$,
except in the outermost regions, where it falls like $\propto R^{-1.4}$
(\S \ref{secc_struc}).

Our calculations rely on the assumption of a steady disk,
of course, which may not be an appropriate assumption.
Nonetheless, detailed analysis of the surface
density distribution of the nearby, and thus bright and
clearly resolved disk around TW Hya indicates that
it is consistent with the predictions of irradiated
disks models (Wilner et al. 2000, 2003, 2005; Qi et al. 2004).

\section {Summary and Conclusions}

We have explored models of irradiated accretion disks where dust settling 
has taken place, modifying the vertical distribution of grain sizes and 
abundances. In these models, dust settling is described in a parametric 
way, using two populations of grains: micron interstellar-like
grains in the upper layers, with a reduced dust to gas mass ratio, 
and mm-sized grains in the lower layers, with an increased dust to gas 
mass ratio. The degree of settling 
is parameterized in terms of the depletion of the upper layers $\epsilon$
(settling increases when $\epsilon$ decreases) 
(see \S 2.2). 

\begin{figure}
\plotone{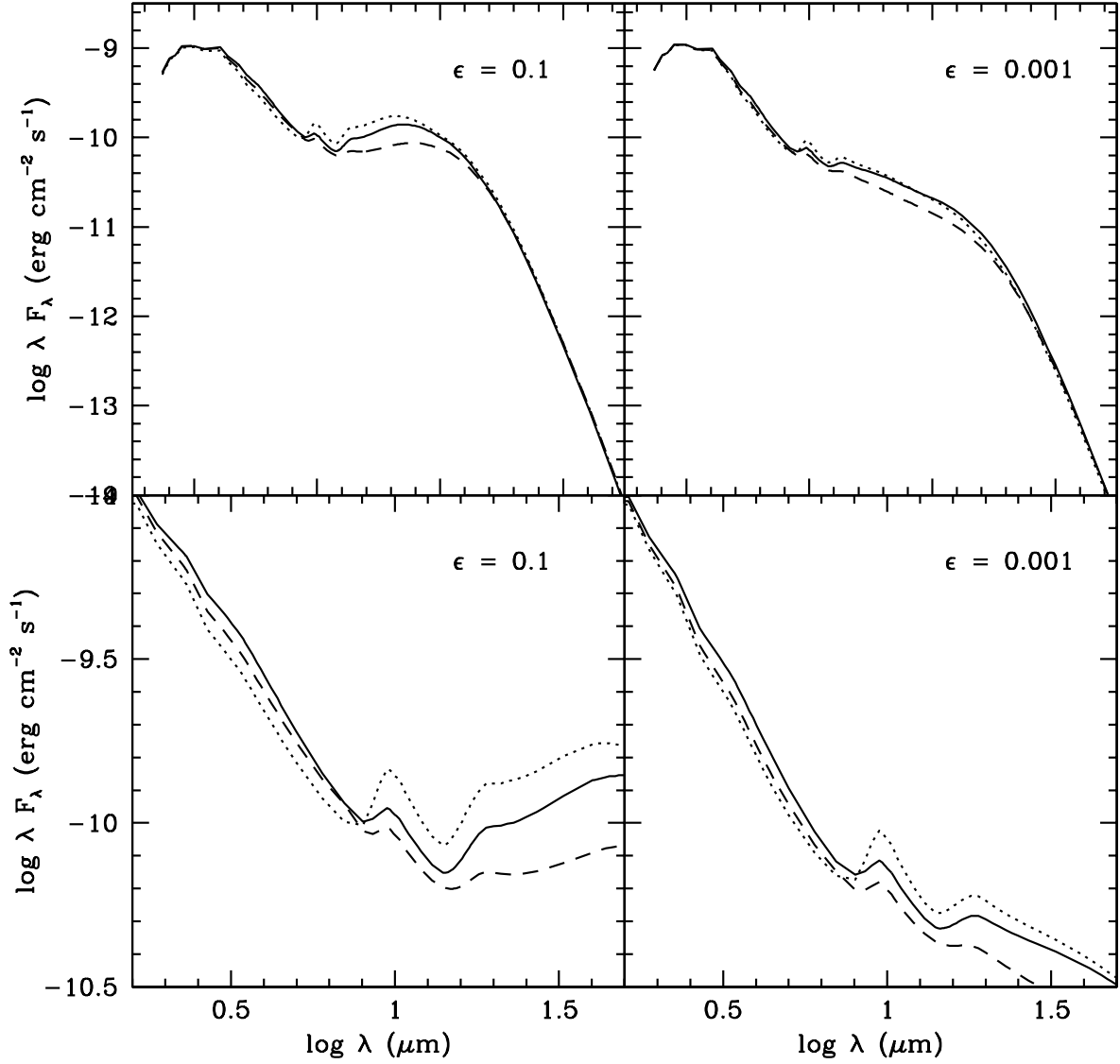}
\caption{SEDs of  the fiducial model  showing the dependence on the
grain size distribution in the upper layers. The models are for
ISM dust and isotropic scattering, for $\epsilon$ = 0.1 (left panels)
and 0.001 (right panels), and for three values is $a_{max}$ for
the dust grains in the upper layers,
$a_{max}$ = 0.05 $\mu$m (dotted lines),
1 $\mu$m (solid lines), and
10 $\mu$m (dashed lines). The lower panels 
show the region of silicate features in more detail.
}
\label{fig_sed_amax}
\end{figure}

\begin{figure}
\plotone{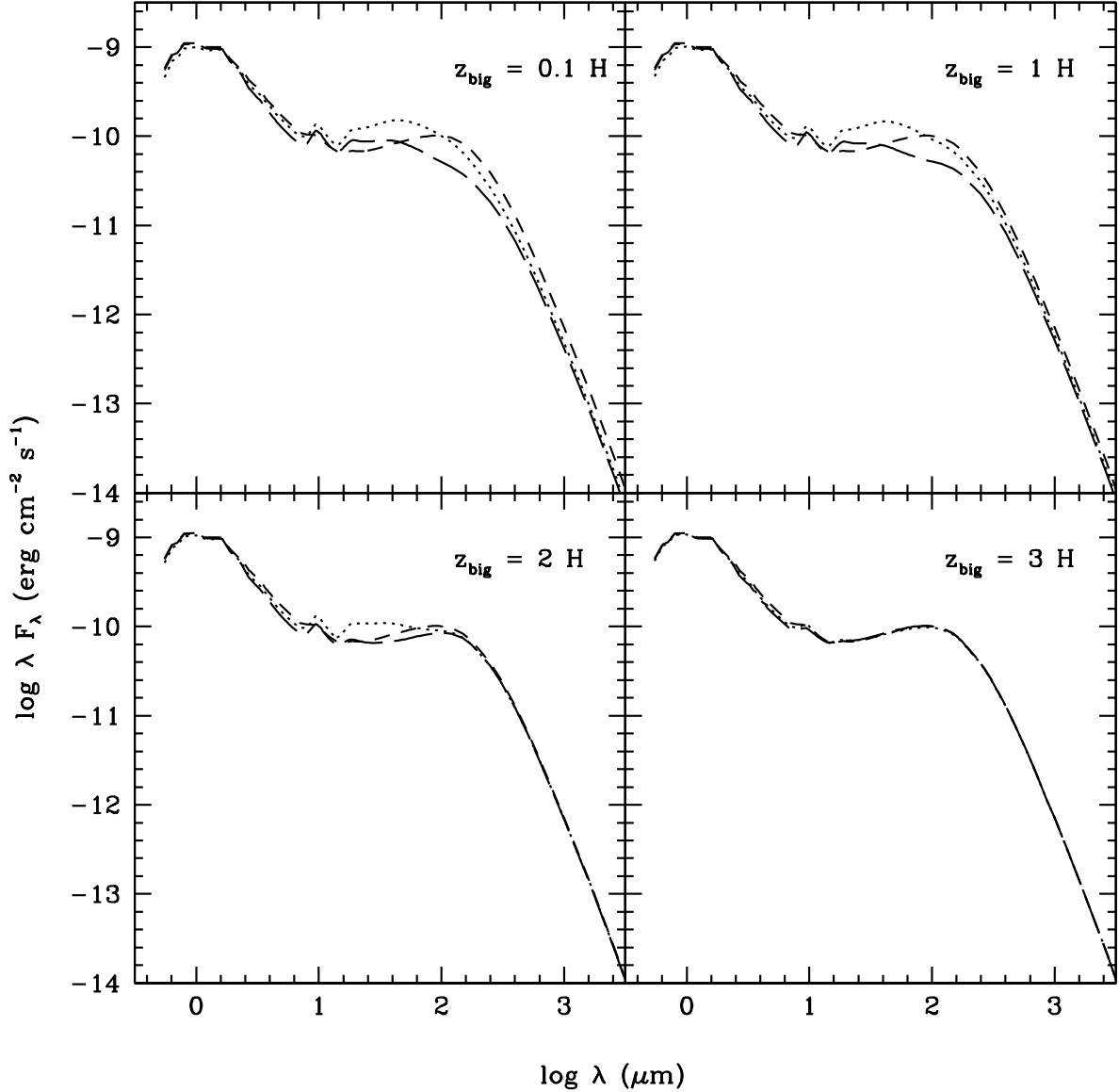}
\caption{SEDs of the fiducial model  showing the dependence on the
height of the big grains layer.
The models are for
ISM dust and isotropic scattering, 
for  $\epsilon=$ 0.1 (dotted line), and 0.01 (dashed line).
and for $z_{big} $ = 0.1, 1, 2, and 3 $H$, where $H$ is the scale
height.
Also the SED of the well mixed model with $a_{max}=$ 1 mm and $\epsilon=$1
is shown in each panel (thick dashed line) for reference.
}
\label{fig_sed_zbig}
\end{figure}

\begin{figure}
\plotone{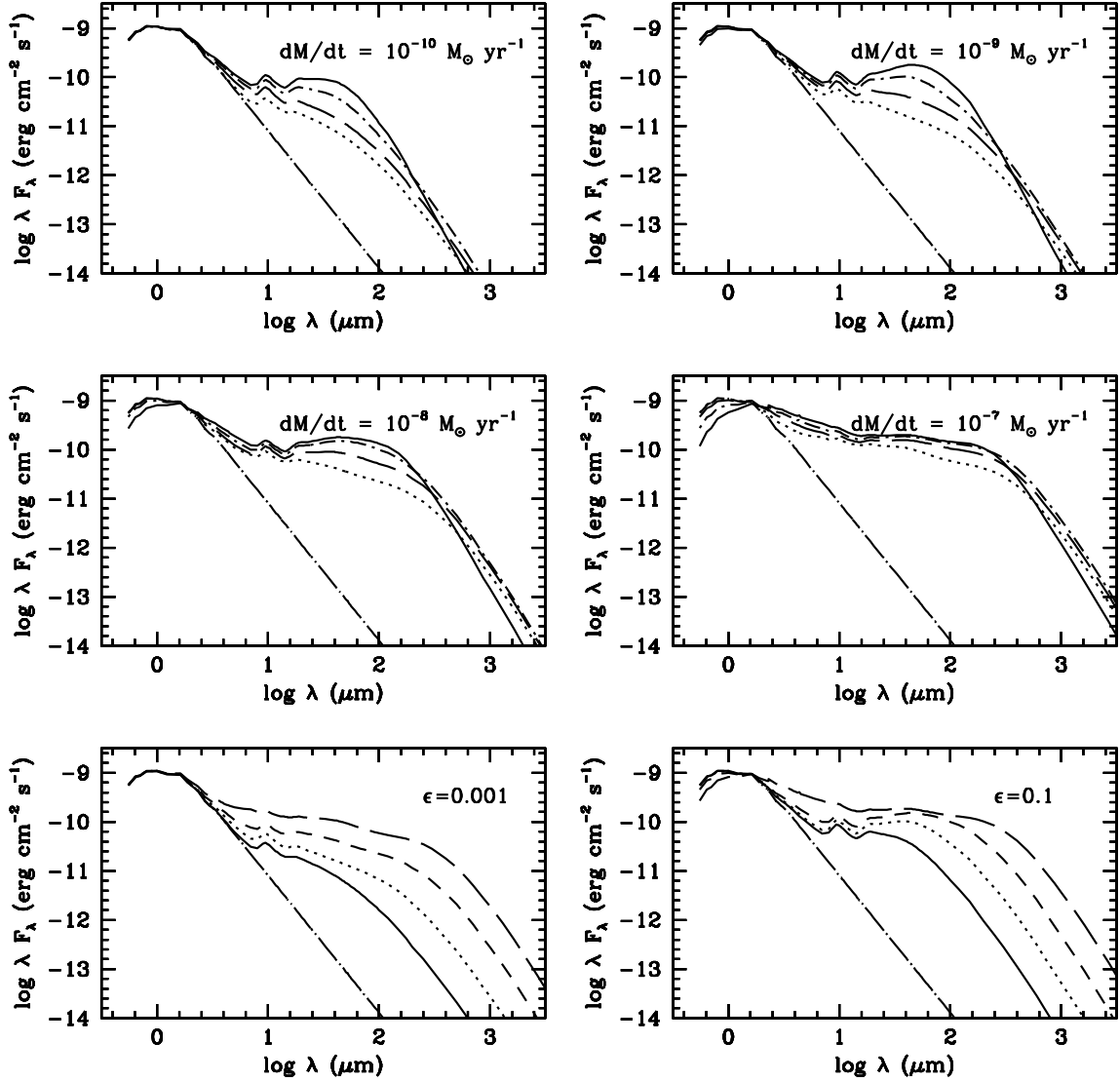}
\caption{SEDs of fiducial disk model with ISM dust, varying the
mass accretion rate.
Each of the four
upper panels correspond each one to a different value of $\Mdot$,
indicated in the panel. Each curve 
correspond to different $\epsilon$:  0.001 (dotted line), 
0.01 (long dashed line), 
0.1 (short dashed line) and 1 (solid line). 
The two lower panels show a selection of the same models for fixed $\epsilon$, 
varying the mass accretion rate: $\Mdot=10^{-10} {\rm (solid)}, 
10^{-9} {\rm (dotted)}, 10^{-8} {\rm (dashed)}$ and $10^{-7} \ \MSUNYR$ (long dashed line) 
}
\label{fig_mdots}
\end{figure}

\begin{figure}
\plotone{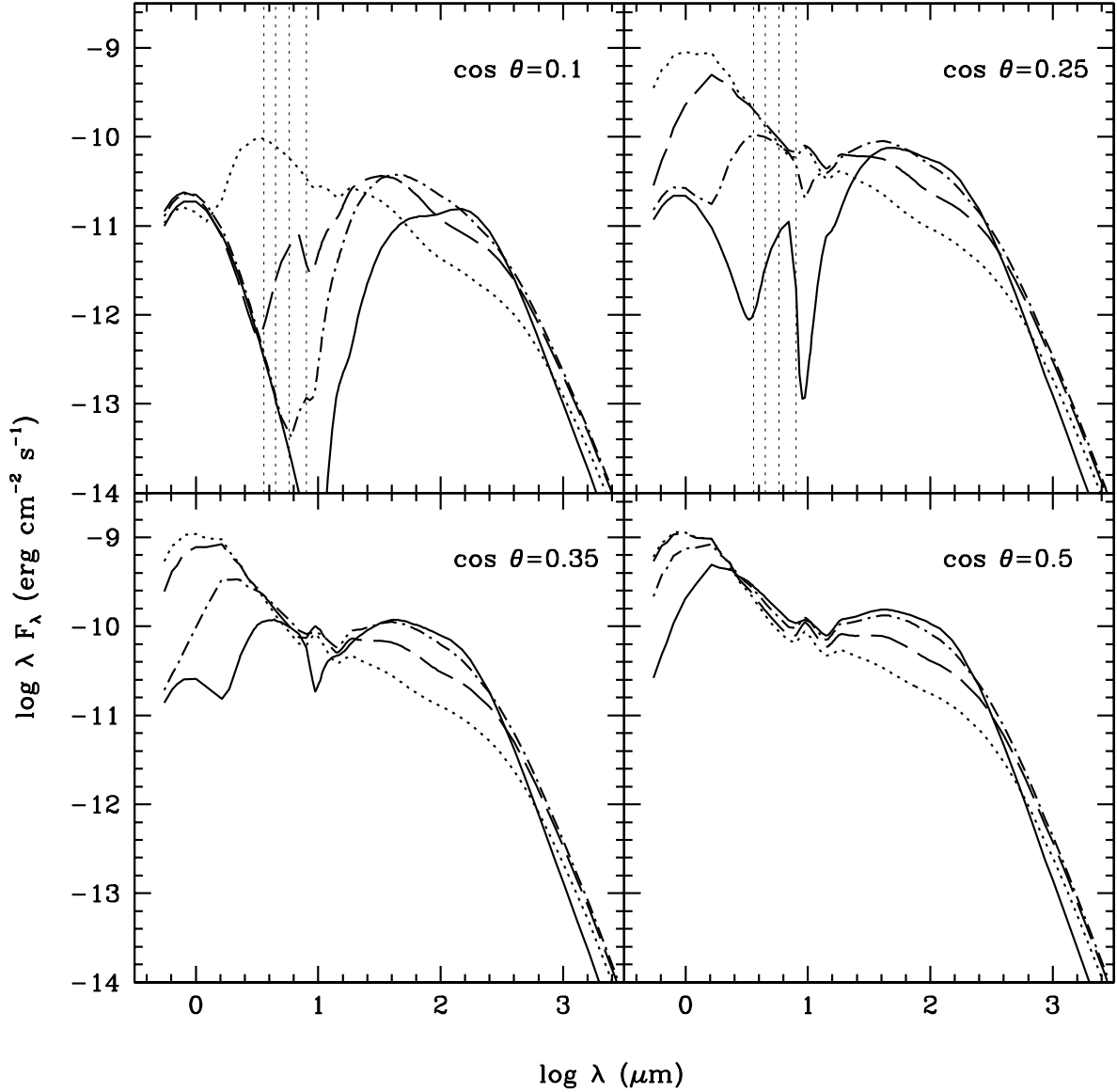}
\caption{SEDs of the fiducial disk model with ISM dust
and isotropic scattering with its axis inclined 
to $\cos i =$ 0.1, 0.25, 0.35 and 0.5 ($i=$ 84, 75.5, 69.5 and 60 $^\circ$)
with respect to the line of sight, and,
four values of the settling parameter:
$\epsilon=$1 (solid), 0.1 (short-dashed),
0.01 (long-dashed) and 0.001 (dotted). 
The dotted lines in the upper panel indicate
the central wavelengths of the IRAC bands.
}
\label{fig_angles}
\end{figure}

\begin{figure}
\plotone{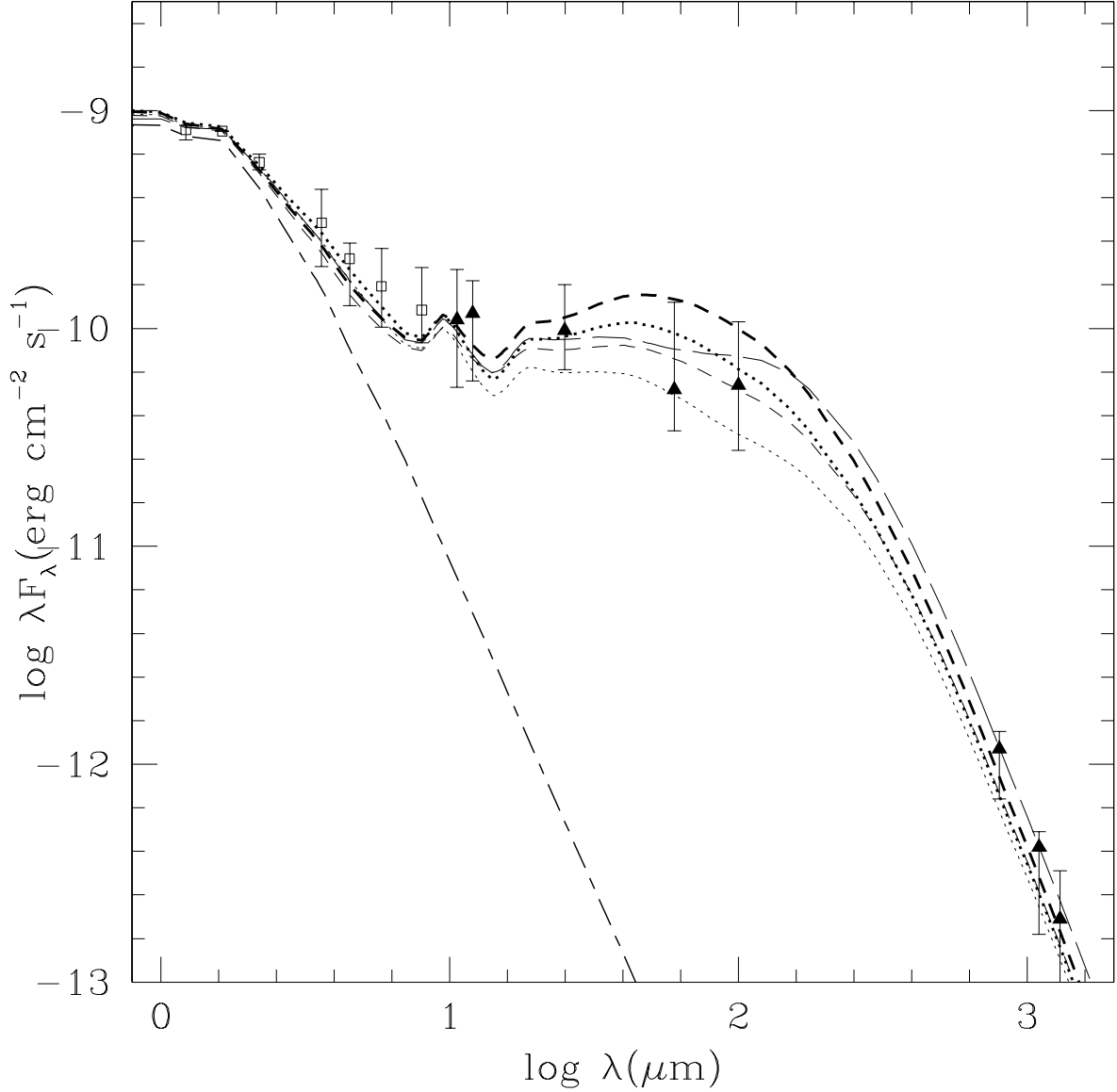}
\caption{SEDs of  the fiducial model with ISM dust, 
for $\epsilon=0.1$ 
(heavy lines) and $\epsilon=0.01$ (light lines),
and for two inclinations: $\mu$ = 0.5 (60$^\circ$, solid lines)
and $\mu$ = 0.87 (30$^\circ$, dashed lines). 
The lower dot-short line indicates the photosphere.
A model with $\epsilon$ = 0.1 and $z_{big}$ = 2 H from
Figure \ref{fig_sed_zbig} is also shown (long-dash line).
The open squares represent the median SED of CTTS
observed with IRAC (Hartmann et al. 2005), the triangles
are the median SED in Taurus from D'Alessio et al. (1999).
In each case,  the error bars correspond to the quartiles. 
}
\label{fig_median}
\end{figure}

\begin{figure}
%\plotone{dalessio_fig16_bw.eps}
\plotone{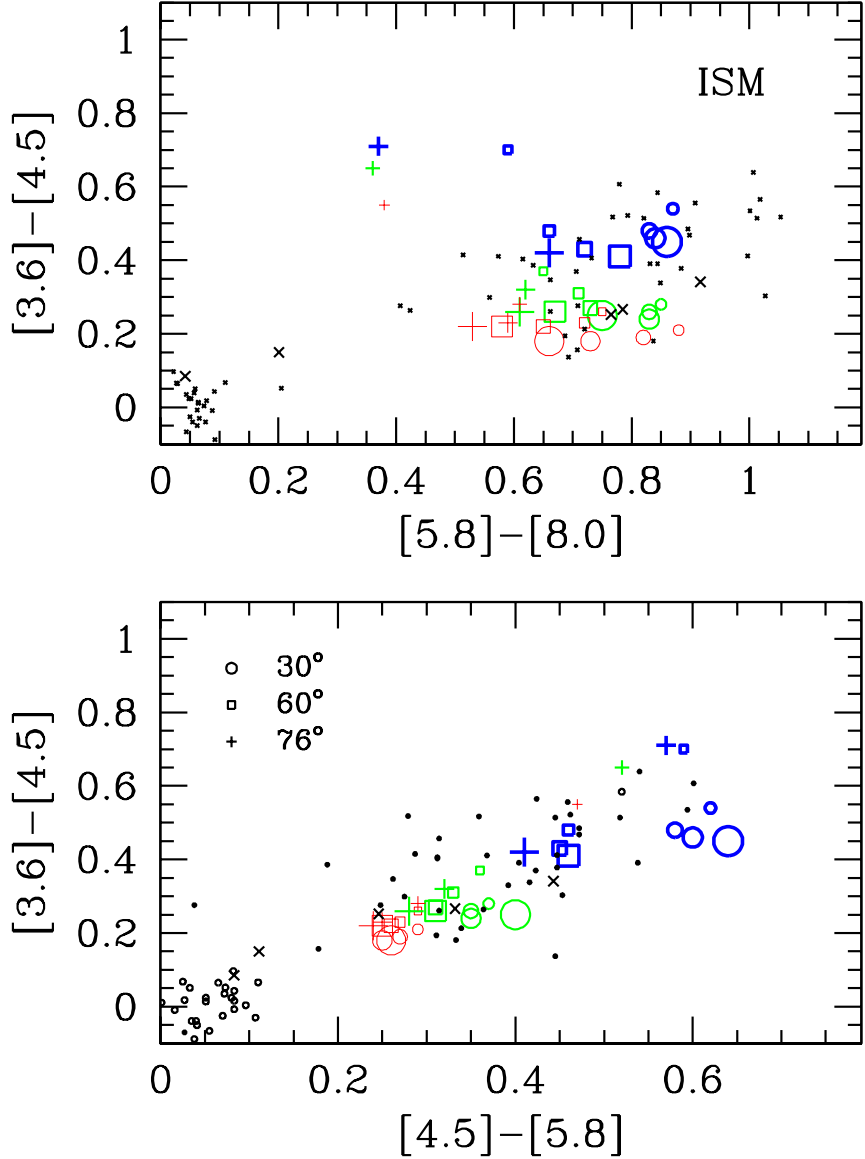}
\caption{
Comparison of predicted and observed IRAC colors
for Classical T Tauri stars in Taurus. The data is taken from Hartmann
et al. (2005) and are shown as 
%black 
small solid dots.
Open small circles are weak T Tauri stars.
Model colors are shown for three mass accretion
rates: 
$10^{-7} \msunyr$ (heavy weight),
%$10^{-7} \msunyr$ (blue),
$10^{-8} \msunyr$ (medium weight), and
%$10^{-8} \msunyr$ (green), and
$10^{-9} \msunyr$ (light weight),
%$10^{-9} \msunyr$ (red),
four values of the settling parameter
$\epsilon = $ 1 (smallest symbols), 0.1, 0.01, and
0.001 (largest symbols),
and three values of inclination
cos$i$ = 0.24 ($i = 76^o$, crosses),
cos$i$ = 0.5 ($i = 60^o$, open squares), and
cos$i$ = 0.87 ($i = 30^o$, open circles).
}
\label{fig_irac}
\end{figure}

\begin{figure}
\epsscale{0.9}
%\plotone{dalessio_fig17_bw.eps}
\plotone{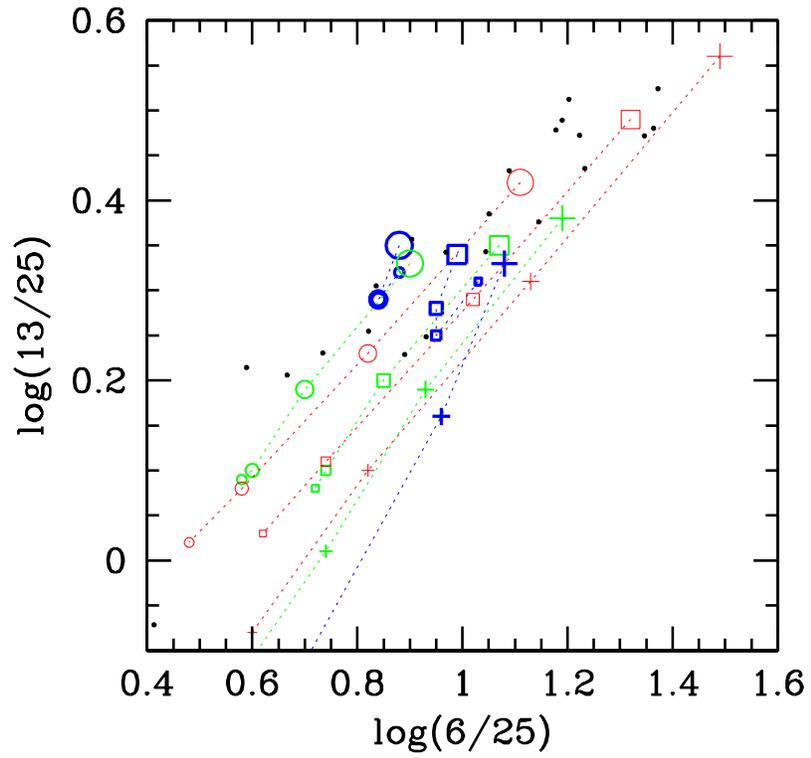}
\caption{Comparison of model predictions with mid-IR
indices derived from IRS spectra for CTTS in Taurus from Furlan et al. (2005a,b).
Observations are shown in black small dots. Model symbols as
in Fig. \ref{fig_irac}. For guidance, dotted lines connect models
with same inclination.
}
\label{fig_irs}
\end{figure}

\begin{figure}
\plotone{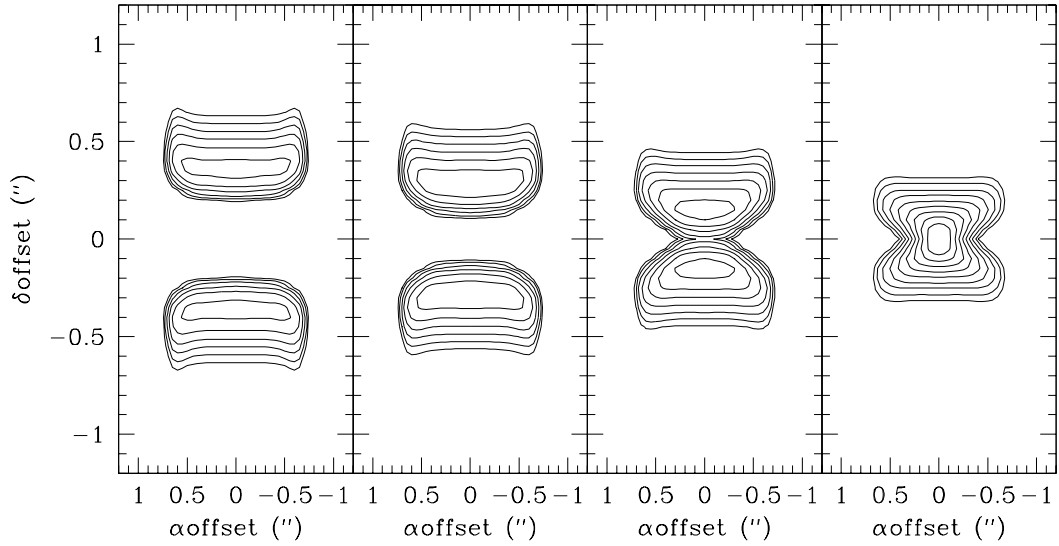}
\caption{Images of edge-on disk models with different depletions at 
$\lambda=0.8 \ \mu m$, convolving with a $0.1\ " $ Gaussian beam. 
>From left to right, 
 $\epsilon=$ 1,  0.1, 0.01,  0.001. The asymmetry factor is 
self-consistently calculated for the dust mixture ($g$ = 0.33).
The plots have the same isocontours, with a maximum
of  0.5 mJy/beam and a minimum of 0.005 mJy/beam.
}
\label{fig_scatt}
\end{figure}

We have found that settling affects the disk structure and SED 
in different ways:

\begin{enumerate} 
\item  The height of the irradiation surface decreases with the degree 
of settling, decreasing the fraction of stellar radiation intercepted 
by the disk and, consequently,  the far IR excess emitted by 
the disk.  The degree of settling affects the degree of flaring of the disk. 
However, even a degree of depletion 
as low as 0.1 \% of the standard 
value, produces a SED with a larger IR excess than a perfectly flat disk, 
reflecting the fact that a very small amount of small dust grains at a few 
scale heights is enough to absorb a larger fraction of stellar radiation 
than predicted by the flat disk approach (\S \ref{secc_emission} and
Fig. \ref{fig_sed}).

\item  The surface temperature of the disk is independent on the degree 
of settling 
but depends on grains sizes and composition. In contrast,
the midplane temperature decreases as 
the degree of settling increases and less stellar radiation is captured
by the disk. 
(Figs. \ref{fig_temperaturas} and \ref{fig_vertical})
This affects the millimeter fluxes emitted by the disk.

\item The mass surface density is as expected for a steady $\alpha$ disk, 
when the viscosity is taken to be a vertical average weighted by mass. 
The resultant radial dependence of the surface density over most of the disk
varies from $\sim R^{-1}$ for no depletion to $\sim R^{-1.2}$
for the most settled models (\S \ref{secc_models} and Fig. \ref{fig_sigma}).
$\Sigma \propto R^{-1.4}$ at large radii where the disk becomes
optically thin.
All the models with $\Mdot=10^{-8} \ \MSUNYR$ 
are gravitationally stable, with the Toomre parameter higher than $\sim 10$.

\item 
The 10 $\mu$m silicate band appears in all the 
models with $a_{max} < 10 \ \mu$m in the upper layers, even in those 
with very small values of 
$\epsilon$ (\S \ref{secc_amax} and Fig. \ref{fig_sed_amax}). 
The band is not affected much by the upper grains dust to gas mass ratio, 
but the continuum is. The intensity of  the band increases
if the size of the grains in the upper layer decrease, because they
hotter (\S \ref{secc_amax} and Fig. \ref{fig_sed_amax}), and 
if the 
scattering of stellar radiation is forward instead of isotropic, because 
the stellar radiation can then penetrate and heat to deeper layers.  

\item
The predicted half-size (region where 50\% of the flux arises) of the
10 micron emission is smaller than 0.5 AU.  For non settled-disks, the
predicted 25 micron half-size can be $\sim$ 10 times larger than the
predicted 10 micron half-size.  However, as the degree of settling
increases, the
half-size of mid-infrared emission decreases, as the optical depth of
the outer disk decreases (\S \ref{secc_spatial} and Figure
\ref{fig_spatial}).

\item Disk models calculated with a degree of settling that changes 
with radius from $\epsilon_1$ to $\epsilon_2$, show SEDs  intermediate 
between two cases with constant depletions (\S \ref{secc_veps}
and Fig. \ref{fig_veps}).

\item 
Disk models calculated with increasing heights of the transition zone 
between both grain populations show SEDs with characteristics of small 
grains at small radii and big grains at larger radii
(\S \ref{secc_zbig} and Fig. \ref{fig_sed_zbig}).

\item Disk models with different mass accretion rates have mid- and far-IR,  
sub-mm and mm fluxes that scale with $\Mdot$ for a given value of $\epsilon$.
 The higher the $\Mdot$, the lower the contrast between the temperatures 
where the continuum and the 10 $\mu$m band forms, and the lower the flux in 
the band relative to the continuum (\S \ref{secc_mdots} and Fig.
\ref{fig_mdots}). 

\item ''Edge-on'' disk models have no direct contribution of emission
from the central star, which is 
being extincted by the outer disk, but show the stellar scattered light 
component which along with the thermal component produces a double bump SED.
As  $\epsilon$ becomes smaller, the disk
becomes flatter and the star less 
extincted. The $10 \mu$m silicate band appears in absorption for the high inclined 
disks, except when $\epsilon$ is very small
(\S \ref{secc_angles} and Fig. \ref{fig_angles}).

\item The median and the observed slopes
in the 3 - 30 $\mu$m range from IRS spectra of CTTS in Taurus 
can be interpreted 
with models with $\epsilon \le 0.1$. A model where the
height of the layer with large grains is $\sim$ 2 scale heights 
reproduce better
the far-IR and millimeter median fluxes of Taurus
(Figs. \ref{fig_median}, \ref{fig_irac}, and \ref{fig_irs}).

\item Comparison with the median IRAC fluxes of Taurus
indicates a deficit of flux in the near-IR and
10 $\mu$m silicate feature when 
compared to models with ISM dust (Fig. \ref{fig_median}). This suggests
that a better treatment of the inner disk region, including
more realistic silicate opacities is necessary.

\item The outermost regions of very settled disks are optically
thin and probably are in the shadow of the inner disk, as indicated
by Dullemond \& Dominik (2004). However, for our disks
where the surface density distribution is self-consistently
calculated, most of the flux
of the very settled disks arises in the inner regions, $< 50 AU$, even at
long wavelengths, so the effects of the shadows  would be difficult
to detect (\S \ref{secc_spatial} and Fig. \ref{fig_spatial}).

\end{enumerate} 

Overall, we have shown that the main features 
in observations of T Tauri disks can be explained 
with models where the upper layers are depleted of dust by 
a factor of ten
or more. We have explored the parameter space covered
by CTTS, and have shown the main effects on the SEDs
of these assumptions. 
Our models admittedly have many simplifying assumptions.
Interpretation of the actual SEDs of individual stars
will provide a better estimate of the extent and degree 
of grain growth and settling, and thus of the degree of dust evolution. 
Efforts to model individual objects with detailed
SEDs will be attempted in forthcoming papers.

\acknowledgments 
PD acknowledges support from Grants from UNAM and CONACyT.
NC and LH acknowledge support from
NASA Grants NAG5-9670, NAG5-13210, and NNG05G126G.

\newpage

\section{Appendix}
\label{secc_append}

Given a power law grain size distribution, the constant $n_0$ is 

\be
\frac{n_0} {n_H}= \frac{3 \zeta m_H \mu (4-p)}  {4 \pi \rho_d (a_{max}^{4-p}-a_{min}^{4-p})},
\en
where $n_H$ is the density of hydrogen nuclei, $m_H$ is the mass of hydrogen, 
$\mu$ is the mean molecular weight relative to hydrogen, and $\rho_d$ is the
bulk density of the grain material.

The mass surface density of dust in the big grains layer is

\be
\Sigma_{down}^d = 2 \zeta_{big} \int_0^{z_{big}} \rho dz
\en
where $\rho$ is the volumetric density of gas. On the other hand, 
the mass surface density of dust that has disappeared from the 
upper disk regions is 
 
\be
\Sigma_{up,lost}^d = 2 \zeta_{std} \biggl [ 1 - \frac{\zeta_{small}}  {\zeta_{std}} \biggr ]  \int_{z_{big}}^\infty  \rho dz
\en
where we assume that the original distribution of dust (before 
settling starts) has a standard 
dust to gas mass ratio, $\zeta_{std}$. In other words, 
the total dust mass surface density is $\Sigma_d = \zeta_{std} \Sigma$, 
where $\Sigma$ is the mass surface density of gas, i.e., 
$\Sigma = 2 \int_0^\infty \rho dz$.

Assuming that all the mass in dust lost by the upper regions is down, 
in the big grains layer, then

\be
\Sigma_{down}^d =2 \zeta_{std} \int_0^{z_{big}} \rho dz + \Sigma_{up,lost}^d
\en

Thus, 

\be
\frac{\zeta_{big}}{\zeta_{std}} = 
\biggl [ \int_0^{z_{big}} \rho dz +  \biggl (1- \frac{\zeta_{small}}  
{\zeta_{std}} \biggr ) 
\int_{z_{big}}^\infty  \rho dz \biggr ] /  \int_0^{z_{big}} \rho dz
\label{eq_exact}
\en

Lets consider $z_{big}$ as a small fraction of the gas scale height, 
i.e., $z_{big} \approx \delta H$, thus the integrals can be approximated 
as following
\be
\int_0^{\delta H} \rho dz \approx \rho_0 \delta H \approx  \frac {\Sigma} {\sqrt{2 \pi} } \delta
\en

\be
\int_{\delta H}^\infty \rho dz \approx    \frac{\Sigma}{2} - \frac{\Sigma} {\sqrt{2 \pi} } \delta  
\en

 Using these approximations (only valid for $\delta << 1$), 
\be
\frac{\zeta_{big}} {\zeta_{std}} = \biggl [ 1 + \biggl (  \frac{\sqrt{2 \pi}}  {2 \delta} -1 \biggr ) \biggl ( 1 - \frac{\zeta_{small}} {\zeta_{std}} \biggr ) \biggr ]
\en

In table \ref {tabla_epsilon}  we list $\zeta_{big}/\zeta_{std}$ for the values of 
$\zeta_{small}/\zeta_{std}$ we use in this paper. 

%\clearpage

\begin{deluxetable}{cc}
\tablecaption{Dust to gas mass ratio \label{tabla_epsilon}}
%\tablewidth{50pt}
\tablehead{
\colhead{$\zeta_{small}/\zeta_{std}$} & \colhead{$\zeta_{big}/\zeta_{std}$} }
\startdata
1 & 1 \\
0.5 & 6.8 \\
0.1 & 11.4 \\
0.05 & 12.0 \\
0.01 & 12.4 \\
0.005 & 12.5 \\
0.001 & 12.5 \\
0.0005 & 12.5 \\
0.0001 & 12.5 
\enddata
\end{deluxetable}
%\clearpage

In order to avoid numerical problems, we make a smooth transition between 
both dust populations such that, 

\be
\zeta_{big}(z) = 0.5 \zeta_{big,0} \biggl \{1 + \tanh\biggl [k\biggl (1-\frac{z} {\delta H} \biggr ) \biggr ] \biggr \}  
\en
\be
\zeta_{small}(z) = 0.5 \zeta_{small,0} \biggl \{1 - \tanh\biggl [k\biggl (1-\frac{z} {\delta H} \biggr ) \biggr ] \biggr \}  
\en
where we have adopted $k=20$, $\delta=0.1$, and the values of 
$\zeta_{big,0}$ and $\zeta_{small,0}$ are those listed in table \ref{tabla_epsilon}.

The midplane layer of big grains has a dust to gas mass ratio larger than
standard, in order to account for the dust that has settled from 
upper layers.

%references
{}

\clearpage

%\end{document}

%figuras

\clearpage

\clearpage

\clearpage

\clearpage

\clearpage

\clearpage

\clearpage

\clearpage

\clearpage

\clearpage

\clearpage

\clearpage

\clearpage

\clearpage

\clearpage

\clearpage

\clearpage

%termina figuras

\end{document}